\documentclass[12pt]{elsart}
\usepackage[cp1251]{inputenc}
\usepackage{graphicx}

\def\be{\begin{eqnarray}}
\def\ee{\end{eqnarray}}
\def\bc{\begin{center}}
\def\ec{\end{center}}
\def\om{\omega}

\def\prt{\partial}
\def\re{{\rm Re}}
\def\im{{\rm Im}}

\def\lsim{\stackrel{\scriptstyle <}{\phantom{}_{\sim}}}
\def\gsim{\stackrel{\scriptstyle >}{\phantom{}_{\sim}}}

\def\rmd{{\rm d}}
\begin{document}
\begin{frontmatter}
\title{Thermodynamics of resonances and blurred particles \\ }
 \author[GSI,MEPhI]{D.N. Voskresensky}
\address[GSI]{GSI, Plankstra\ss{}e 1, D-64291 Darmstadt, Germany}
\address[MEPhI]{Moscow Engineering Physical Institute,\\ Kashirskoe
  Avenue 31, RU-115409 Moscow, Russia}

\begin{abstract}
Exact and approximate expressions for thermodynamic
characteristics of heated matter, which consists of
 particles with finite mass-widths,  are constructed.
They are  expressed in terms
 of  Fermi/Bose distributions and
 spectral functions, rather than in terms of  more complicated combinations between real
 and imaginary parts of the self-energies of different particle species.
Therefore thermodynamically consistent approximate treatment of
systems of particles with finite mass-widths  can be performed,
provided spectral functions of particle species are known.
Approximation of the free resonance gas at low densities is
studied. Simple ansatz for the energy dependence of the spectral
function  is suggested that allows to fulfill  thermodynamical
consistency conditions. On examples it is shown that a simple
description of dense systems of interacting particle species can
be constructed, provided some species can be treated in the
quasiparticle approximation and others as particles with widths.
The interaction affects quasiparticle contributions, whereas
particles with widths can be treated as free. Example is
considered of a hot gas of heavy fermions strongly interacting
with light bosons, both species with zero chemical potentials. The
density of blurred fermions is dramatically increased for high
temperatures compared to the standard Boltzmann value. The system
consists of boson quasiparticles (with effective masses)
interacting with fermion-antifermion blurs. In thermodynamical
values interaction terms partially compensate each other. Thereby,
in case of a very strong coupling between species thermodynamical
quantities of the system, like the energy, pressure and entropy,
prove to be such as for the quasi-ideal gas mixture of quasi-free
fermion blurs and quasi-free bosons.

\end{abstract}\end{frontmatter}

\section{Introduction}
In heavy ion collisions at relativistic energies baryon resonances
may play an important role in the reaction dynamics \cite{SCFNW}.
Therefore the name "Resonance matter" has been coined for highly
excited matter, where a large fraction of nucleons is converted to
resonance states. We should specify terms. One deals with a
resonance, if the spectral function of the particle has rather
sharp peak as function of the energy. In spite of the imaginary
part of the self-energy is finite, the pole-like structure of the
Green function still remains.
 In many models
resonances are however treated as quasiparticles. The
quasiparticle approximation means that one may put imaginary part
of the self-energy zero in the expression for the Green function.
Then the spectral function becomes  the delta-function. Definitely
quasiparticle approximation for resonances is only a rough
approximation being done  for simplification. E.g., in vacuum the
$\rho$-meson width at the maximum is as high as about $150$~MeV
and the $\Delta$-resonance width is  about $120$~MeV.
 Resonances with masses up
to 2 GeV at RHIC and LHC energies contribute to thermodynamics of
the fireball \cite{BMRS}. Some of them have still larger widths.
In medium  particle widths may get an extra increase due to
collision broadening. It may result in a complete blurring of
particles. By blurring of the particle we understand here the
case, when the Green function is entirely regular. Blurring of
fermions may happen, e.g., at sufficiently high temperature and a
small baryon concentration \cite{Dyug,V04}. Exact solution of the
problem is not possible, even if one restricts consideration by a
specific choice of the interaction and considers only few
particle species. Therefore it is a challenge to construct a
simplified  but thermodynamically consistent  description of
heated matter consisting of quasiparticles, resonances and blurred
particles.

In this paper we present exact and approximate expressions for
thermodynamic characteristics of heated  resonance matter that
consists of
 particles with finite mass-widths (resonances and blurs).
 Thermodynamic quantities are  expressed in terms
 of  Fermi/Bose distributions and
 spectral functions, rather than in terms of  more complicated combinations between real
 and imaginary parts of the self-energies of different particle species, see sect. \ref{Thermodynamic}.
 Solution of the problem is achieved provided  spectral
 functions of  particles with widths are known. Different examples,
 when such a description might be helpful are presented. In sect.
 \ref{Gas} we consider gas of free resonances (at low densities).
 Then in sect. \ref{System} we formulate how one could proceed in
 description of dense systems of strongly
 interacting particles with widths in cases, when some species can be treated as quasi-free broad resonances
 or blurred particles,
 whereas other species, as interacting quasiparticles.  Different examples are considered.
 In sect. \ref{Quasi} we study example of a strongly interacting hot heavy fermion
 -- light boson system at zero chemical potentials of species.
 Concluding remarks are presented in sect. \ref{Concluding}.

\section{Thermodynamic
quantities in terms of spectral functions} \label{Thermodynamic}

\subsection{General  relations}

 The thermodynamic potential density $\Omega$, pressure $P$, free
 energy density $F$, energy density $E$ and entropy density $S$
 follow thermodynamic
 relations:
 \be\label{presmu} 
E=F+TS, \quad
  F[f,\om_0 ,R_0 ] =\sum_{i} \mu_i n_i +\Omega\,,\quad \Omega =-P,
\,
  \ee
where $T$ is the temperature and \be\mu_i=\frac{\prt F}{\prt
n_i}\ee is the chemical potential,  and $n_i$ is the density of
the $i$-particle species.

The density and the entropy density are the derivatives of the
pressure:
 \be\label{consist}
 n_{i} =\frac{\partial P_{i}}{\partial
\mu_{i}}|_{T}, \quad S_i =\frac{\partial P_i}{\partial
T}|_{\mu_{i}}.\ee
These conditions should be fulfilled  at any
thermodynamically consistent description of  properties of the
medium.

It is convenient to introduce one-particle spectral and width
functions (operators), cf. \cite{IKV2,V04},
 \be\label{A-G}
\widehat{A} =-2\im \widehat{G}^R (q)=-2\im\frac{1} { \widehat{M}
+i\widehat{\Gamma} /2},\,\,\, \widehat{\Gamma} =-2\im
\widehat{\Sigma}^R \,,
 \ee
where $\widehat{G}^R (q)$ is the full retarded Green function of
the resonance, $\widehat{\Sigma}^R$ is the fermion retarded
self-energy. The quantity
 \be\label{M}
\widehat{M}=(\widehat{G}^{0,R})^{-1}-\re \widehat{\Sigma}^R
 \ee
demonstrates  deviation from the mass shell: $\widehat{M}=0$ on
the quasiparticle mass shell in the matter. $\widehat{G}^{0,R}$ is
the free retarded Green function.

\subsection{Non-relativistic particles}\label{Spectral}

For non-relativistic particles (fermions or bosons) spectral
function satisfies the sum rule:
 \be\label{sex}
\int_{-\infty}^{\infty}A \frac{d
  \om}{2\pi}=1.
   \ee
Note that although  in (\ref{sex}) the lower integration limit is
taken to be $-\infty$, the spectral function drops to zero at some
finite threshold value $\om_{\rm th}$. Therefore in reality the
lower integration limit is $\om_{\rm th}$.

In the quasiparticle limit, when the width $\Gamma$ is much less
than all other relevant physical quantities, the spectral function
acquires a $\delta$ -function shape
 \be
A\rightarrow A^{\rm q.p.}_{\rm n.r.}=2\pi \delta (\om -\om_p^{\rm
n.r.}-\re\Sigma^R_{\rm n.r.} (\om, {p})),
 \ee
$\om_p^{\rm n.r.} =p^2 /(2m)$, $m$ is the particle mass.  The sum
rule renders
 \be
\int_{-\infty}^{\infty}A^{\rm q.p.}_{\rm n.r.} \frac{d
  \om}{2\pi}= 1-\frac{\partial
  \mbox{Re}\Sigma^R_{\rm n.r.}}{\partial \om}\mid_{\om (p)}>0,
   \ee
where $\om (p)= \om_p^{\rm n.r.} +\mbox{Re}\Sigma^R_{\rm n.r.}
\left(\om (p), {p}\right)$. To fulfill the exact sum-rule
(\ref{sex}) one should take into account a compensating
contribution of the whole sea of off-shell modes.

A self-consistent description of non-equilibrium and equilibrium
resonance matter beyond the scope of the quasiparticle
approximation can be constructed basing on solution of
Kadanoff-Baym equations within a $\Phi$ derivable approach, cf.
\cite{IKV1,IKV2,IKV3}.
 Refs. \cite{IKV1,IKV2} have derived exact Noether
expressions for the particle 4-current and the energy-momentum
tensor. Results are formulated for non-relativistic particles
\cite{IKV2} and for relativistic bosons \cite{IKV1} in case of
interactions with non-derivative couplings.
\footnote{Generalizations to  systems with derivative couplings
are done in \cite{IKV3}.} For non-relativistic particles:
 \be\label{Noet}j^{\mu}_{\rm n.r.}=\mbox{Tr}
\int\frac{d^4 p}{(2\pi)^4} v^{\mu} A_{\rm n.r.} f,
 \ee
 \be\label{energymom}\Theta^{\mu\nu}_{\rm n.r.}=\mbox{Tr}
\int\frac{d^4 p}{(2\pi)^4}v^{\mu}p^{\nu} A_{\rm n.r.} f
+g^{\mu\nu}(\epsilon^{\rm int}_{\rm n.r.}-\epsilon^{\rm pot}_{\rm
n.r.}),
 \ee
 with $v^{\mu}=(1, \vec{v})$, $\vec{v}=\vec{p}/m$. Particle occupations are given by
 \be f=\frac{1}{e^{(\om -\mu_{\rm n.r.} )/T}\pm 1},
  \ee
with "$+$" for fermions, "$-$" for bosons; for a nucleon resonance
such as $\Delta$ from equilibrium condition in respect to the
reaction $\Delta \leftrightarrow N+\pi$ it follows that  $
\mu^{\rm n.r.}_{\Delta} =\mu_{N}^{\rm n.r.}$ (since
$\mu_{\pi}=0$); $\epsilon^{\rm int}$ and $ \epsilon^{\rm pot}$ are
some interaction and potential energies. Explicit expressions for
them are presented below. Here and below, considering a hot
system,  we disregard a contribution of quantum fluctuations.

From (\ref{energymom}) we find the energy density
 \be\label{enrel}
\Theta^{00}_{\rm n.r.}=E_{\rm n.r.} =N_{\rm n.r.} \int\frac{d^4
p}{(2\pi)^4} \om A_{\rm n.r.} f + \epsilon^{\rm int}_{\rm
n.r.}-\epsilon^{\rm pot}_{\rm n.r.},
 \ee
where $N_{\rm n.r.}$ is the  degeneracy factor, that appears as
the result of taking the trace in (\ref{energymom}), and the
pressure
 \be\label{pnrel} P_{\rm
n.r.}&=&\frac{1}{3}(\Theta^{11}+\Theta^{22}+\Theta^{33})\nonumber\\&=&N_{\rm
n.r.} \int\frac{d^4 p}{(2\pi)^4} \frac{p^2}{3m} A_{\rm n.r.} f -
\epsilon^{\rm int}_{\rm n.r.}+\epsilon^{\rm pot}_{\rm n.r.}.
 \ee
Using these expressions and also Eq. (\ref{Noet}), from
thermodynamical relation (\ref{presmu}), Ref. \cite{IKV2} derived
{\em{exact expression for the entropy density}}:
 \be\label{stnon}
T S_{\rm n.r.}=N_{\rm n.r.} \int\frac{d^4 p}{(2\pi)^4}(\om
+\frac{2}{3}\om_p^{\rm n.r.} -\mu_{\rm n.r.}) A_{\rm n.r.} f .
 \ee
The interaction terms $\epsilon^{\rm int}$ and $ \epsilon^{\rm
pot}$ have cancelled due to  presence of the $g^{\mu\nu}$ factor
in (\ref{energymom}).

Note that both expressions (\ref{Noet}) and (\ref{stnon}) for the
4-current and for the entropy are exact and both quantities depend
on the specifics of the interaction only through the $A$-spectral
function. On the other hand,  both the energy (\ref{enrel}) and
the pressure (\ref{pnrel}) depend also on a combination
$\epsilon^{\rm int}-\epsilon^{\rm pot}$ of interaction and
potential energies.

\subsection{Relativistic bosons}

For relativistic spin-less charged bosons\footnote{By the charge
we mean any conserved quantity like electric charge, strangeness,
etc.} the spectral function can be represented in terms of
spectral functions of particles and antiparticles, cf.
\cite{IKV3},
 \be
A_{\rm b}(\om , \vec{p}\,)=A_{\rm b}^{(+)} (\om , \vec{p}\,)\theta
(\om)-A_{\rm b}^{(-)} (-\om , -\vec{p}\,)\theta (-\om),\ee
\be
A_{\rm b}^{(+)}(\om , \vec{p}\,)=A_{\rm b} (\om , \vec{p}\,)\theta
(\om), \quad A_{\rm b}^{(-)}(\om , \vec{p}\,)=-A_{\rm b} (-\om ,
-\vec{p}\,)\theta (\om),
 \ee
 \be
\Sigma^R_{\rm b}(\om , \vec{p}\,)=\Sigma^{R{(+)}}_{\rm b} (\om ,
\vec{p}\,)\theta (\om)+ \Sigma^{A{(-)}}_{\rm b} (-\om ,
-\vec{p}\,)\theta (-\om).
 \ee
 Here $\theta (x)$ is the
step-function, "$R$" denotes retarded and "$A$", advanced
quantity.

 The sum rule renders:
  \be
\int_{0}^{\infty}\om\frac{ d \om}{2\pi}\left[A_{\rm b}^{(+)}(\om ,
\vec{p}\,) +A_{\rm b}^{(-)}(\om , -\vec{p}\,)\right] =1.
 \ee
In the quasiparticle approximation the spectral function becomes
 \be\label{qpbos}
 [A^{\rm q.p.}_{\rm b}]^{(\pm)}=2\pi \delta \left[\om^2 -(\om_p^{\rm
b})^2 -\re\Sigma^{R{\rm (\pm)}}_{\rm b} (\om, {p}\,)\right],
 \ee
with $[\om_p^{\rm b}]^2 =m_{\rm b}^2 + {p}^{\,\,2},$
 and
the sum rule reads
 \be
\int_{0}^{\infty}[A^{\rm q.p.}_{\rm b}]^{(\pm)} \om \frac{d
  \om}{2\pi}=\frac{1}{2} \left[1-\frac{\partial
  \mbox{Re}\Sigma^{R(\pm)}_{\rm b}}{\partial \om^2}\mid_{\om (p)}\right]>0,
   \ee
where $\om^2 (p) =m_{\rm b}^2 + {p}^{\,2} +\re\Sigma^{R(\pm)}_{\rm
b} \left(\om (p), {p}\right)$.

 Expression for the energy-momentum
tensor of the boson sub-system looks similar to Eq.
(\ref{energymom}), see \cite{IKV1},
 \be\label{energymomB}\Theta^{\mu\nu}_{\rm
b}=
\int\frac{d^4 p}{(2\pi)^4}2p^{\mu}p^{\nu} A_{\rm b}
f_{\rm b }+g^{\mu\nu}(\epsilon^{\rm int}_{\rm b}-\epsilon^{\rm
pot}_{\rm b}),
 \ee
where now
  \be f_{\rm b}=\frac{1}{e^{(\om -\mu_{\rm b})/T}-1}
   \ee
are boson occupations, and $\mu_{\rm b}$ is the boson chemical
potential. Expressions for interaction and potential energies,
$\epsilon^{\rm int}_{\rm b}$ and $ \epsilon^{\rm pot}_{\rm b}$,
are presented below.

Refs. \cite{IKV1,IKV3} also obtained {\em{exact}} Noether
expressions for the charged boson density, the energy density and
the pressure:
 \be\label{boso} n_{\rm b}=
\int_{0}^{\infty}\frac{4\pi p^2 d
p}{(2\pi)^3}\int_{0}^{\infty}\frac{ d \om}{2\pi} 2\om A_{\rm
b}^{(+)} f_{\rm b}^{(+)} -(\mu_{\rm b}\rightarrow -\mu_{\rm b}) ,
 \ee
 \be\label{Ebo}
 E_{\rm b}=
\int_{0}^{\infty}\frac{4\pi p^2 d
p}{(2\pi)^3}\int_{0}^{\infty}\frac{ d \om}{2\pi}2\om^2 A_{\rm
b}^{(+)} f_{\rm b }^{(+)}+\epsilon_{\rm int}^{{\rm
b}(+)}-\epsilon_{\rm pot}^{{\rm b}(+)}+(\mu_{\rm b}\rightarrow
-\mu_{\rm b}),
 \ee
 \be\label{Pbo} P_{\rm b}=
\int_{0}^{\infty}\frac{4\pi p^2 d
p}{(2\pi)^3}\int_{0}^{\infty}\frac{ d \om}{2\pi}\frac{2{p}^2
}{3}A_{\rm b}^{(+)} f_{\rm b}^{(+)} -\epsilon_{\rm int}^{{\rm
b}(+)}+\epsilon_{\rm pot}^{{\rm b}(+)}+(\mu_{\rm b}\rightarrow
-\mu_{\rm b}),
 \ee
 \be\label{stb} T S_{\rm b}=
\int_{0}^{\infty}\frac{4\pi p^2 d
p}{(2\pi)^3}\int_{0}^{\infty}\frac{ d \om}{2\pi} 2\om (\om
+\frac{{p}^{\,2}}{3\om}-\mu_{\rm b}) A_{\rm b}^{(+)} f_{\rm b
}^{(+)}+(\mu_{\rm b}\rightarrow -\mu_{\rm b}).
 \ee
 Antiparticle contributions are included with the help of the replacement $\mu_{\rm b}\rightarrow -\mu_{\rm
b}$.

\subsection{Relativistic fermions}

For spin $1/2$ fermion, spectral function satisfies the sum rule:
 \be\label{fsum-r} \frac{1}{4} \mbox{Tr} \int^{\infty}_0 \gamma_0
\left[\widehat{A}_{\rm f }^{(+)} (\om , \vec{p})+ \widehat{A}_{\rm
f}^{ (-)} (\om ,-\vec{p})\right] \frac{d
  \om}{2\pi}=2,\,\,\,
 \ee
 $\gamma_0$ is the  Dirac matrix.
 The trace is taken over spin degrees of freedom.

In  cases when spin degrees of freedom decouple it is sufficient
to introduce two functions (for each particle species). E.g.,
 \be\label{an}  \widehat{A}_{\rm f}^{
(+)}=\widehat{p}\widetilde{A}_{\rm f}+A_{{\rm f}(1)},\quad
\frac{1}{4}\mbox{Tr}\gamma_0 \widehat{A}_{\rm f}^{(+)} =A_{\rm
f}=\widetilde{A}_{\rm f}\om ,
 \ee
 where we explicitly separated
$\om$-pre-factor. Then generalizations of non-relativistic
expressions for thermodynamic quantities
   to the case of
relativistic fermions are simple. From (\ref{Noet}),
 (\ref{energymom}) we recover expressions for the baryon density, the
 energy density, the pressure and the entropy density:
 \be\label{nd} n_{\rm f}=
 N_{\rm
f}\int_{0}^{\infty}\frac{4\pi p^2 d
p}{(2\pi)^3}\int_{0}^{\infty}\frac{ d \om}{2\pi} A_{\rm f}^{(+)}
f_{\rm f}^{(+)}-(\mu_{\rm f}\rightarrow -\mu_{\rm f}),
 \ee
 \be\label{ed} E_{\rm f}=
N_{\rm f}\int_{0}^{\infty}\frac{4\pi p^2 d
p}{(2\pi)^3}\int_{0}^{\infty}\frac{ d \om}{2\pi}\om A_{\rm
f}^{(+)} f_{\rm f }^{(+)}+\epsilon_{\rm int}^{{\rm
f}(+)}-\epsilon_{\rm pot}^{{\rm f}(+)}+(\mu_{\rm f}\rightarrow
-\mu_{\rm f}),
 \ee
 \be\label{p} P_{\rm f}=N_{\rm
f}\mbox{Tr}\int_{0}^{\infty}\frac{4\pi p^2 d
p}{(2\pi)^3}\int_{0}^{\infty}\frac{ d
\om}{2\pi}\frac{{p}^2}{3\om}A_{\rm f}^{(+)} f_{\rm f}^{(+)}
-\epsilon_{\rm int}^{{\rm f}(+)}+\epsilon_{\rm pot}^{{\rm
f}(+)}+(\mu_{\rm f}\rightarrow -\mu_{\rm f}),
 \ee
 \be\label{st} T S_{\rm f}=N_{\rm f}\int_{0}^{\infty}\frac{4\pi p^2 d
p}{(2\pi)^3}\int_{0}^{\infty}\frac{ d \om}{2\pi} (\om
+\frac{{p}^2}{3\om}-\mu_{\rm f}) A_{\rm f}^{(+)} f_{\rm
f}^{(+)}+(\mu_{\rm f}\rightarrow -\mu_{\rm f}),
 \ee
 \be f_{\rm f}^{(+)}=\frac{1}{e^{(\om -\mu_{\rm f})/T}+1},
  \ee
$N_{\rm f}$ is the  degeneracy factor. The  replacement $(\mu_{\rm
f}\rightarrow -\mu_{\rm f})$ takes into account antiparticle
terms.

\subsection{Relations between interaction and potential energies}

We will use a relation \cite{IKV1,IKV2} between $\epsilon_{\rm
int}$ and $\epsilon_{\rm pot}$:
 \be\label{intpotrel}\epsilon_{\rm
int}=\frac{2}{\alpha}\epsilon_{\rm pot}
 \ee
 for
specific interactions with a certain number $\alpha$ of operators
attached to the vertex. This relation is derived using the
operator forms of $\widehat{\epsilon}_{\rm int}$ and
$\widehat{\epsilon}_{\rm pot}$:
 \be
&&\widehat{\epsilon}_{\rm pot}=\sum_{i}\widehat{\epsilon}^{\rm
pot}_i =-\frac{1}{2}\sum_{i}(\widehat{J}^{\dagger}_i
\widehat{\phi}_i +\widehat{J}_i
\widehat{\phi}_i^{\dagger}),\nonumber\\
&&G_0^{-1}\widehat{\phi}_i=-\widehat{J}_i =-\frac{\delta
\widehat{L}^{\rm int}}{\delta\widehat{\phi}_i^{\dagger}},
\quad\widehat{\epsilon}_{\rm int}=-\widehat{L}^{\rm int},
 \ee
where $\widehat{L}^{\rm int}$ is the interaction term in the
Lagrangian density.

 For two-body non-relativistic interaction and for relativistic boson
 $\phi^4$
theory one gets $\alpha =4$. For a theory with two single-flavor
fermions interacting via one-flavor boson (with coupling
$\Psi_{\rm f}^{\dagger}\Psi_{\rm f} (\phi_{\rm b} +\phi_{\rm
b}^{\dagger})$) one obtains
 \be\label{twoFoneB}\epsilon_{\rm int}=\frac{2}{\alpha}(\epsilon_{\rm
pot}^{\rm f} +\epsilon_{\rm pot}^{\rm b} )=\frac{2}{\alpha_{\rm
f}} \epsilon_{\rm pot}^{\rm f} =\frac{2}{\alpha_{\rm b}}
\epsilon_{\rm pot}^{\rm b} , \quad \alpha =3, \,\, \alpha_{\rm
f}=2,\,\,\alpha_{\rm b}=1.
 \ee
For a theory where two  fermions with different flavors interact
via one-flavor boson, one finds
 \be\label{twodFoneB}\epsilon_{\rm int}=2\epsilon_{\rm pot}^{{\rm f}_1}=2\epsilon_{\rm pot}^{{\rm f}_2}
=2\epsilon_{\rm pot}^{\rm b} .
 \ee

We will also use {\em{exact}} expressions of Refs.
\cite{IKV2,IKV3} for $\epsilon_{\rm pot}$, which follow directly
from equations of motion:
 \be\label{pair} \epsilon_{\rm pot}^{\rm
n.r.}= N_{\rm n.r.}\int_{0}^{\infty}\frac{4\pi p^2 d
p}{(2\pi)^3}\int_{-\infty}^{\infty}\frac{ d \om}{2\pi}(\om
-\om_p^{\rm n.r.} )A_{\rm n.r.}f
 ,
  \ee
  with $\om_p^{\rm n.r.} =p^2 /2m $ for
non-relativistic particles  of given species;

 \be\label{f-b} \epsilon_{\rm pot}^{{\rm b}}=
 \int_{0}^{\infty}\frac{4\pi p^2 d
p}{(2\pi)^3}\int_{0}^{\infty}\frac{ d \om}{2\pi}\left[\om^2_{\rm
b} -(\om_p^{\rm b})^2 \right]A_{\rm b}^{(+)} f_{\rm b}^{(+)}
+(\mu_{\rm b}\rightarrow -\mu_{\rm b}) ,
  \ee
with $\om_p^{\rm b} =\sqrt{p^2 +m^2_{\rm b}}$ for relativistic
spin-less bosons of given species; and
 \be\label{pairR}
\epsilon_{\rm pot}^{{\rm f}}= N_{\rm f}\int_{0}^{\infty}\frac{4\pi
p^2 d p}{(2\pi)^3}\int_{0}^{\infty}\frac{ d \om}{2\pi}(\om_{\rm f}
-\om_p^{\rm f} )A_{\rm f}^{(+)}f_{\rm f}^{(+)} +(\mu_{\rm
f}\rightarrow -\mu_{\rm f})
 ,
  \ee
 with
$\om_p^{\rm f} =\sqrt{p^2 +m^2_{\rm f} }$ for relativistic
fermions  of the given species (for interactions with  a simple
spin structure of vertices).

Eqs. (\ref{pair}), (\ref{f-b}), (\ref{pairR}) demonstrate that in
general case
 even at low densities the potential
energy may not cease.

Convenience of expressions for thermodynamic characteristics of
the system that we have presented is that all quantities are
expressed entirely in terms of spectral functions.  Thus the
problem of the description of strongly interacting matter would be
solved, if  spectral functions of all species were known and if
there were relations between all partial interaction and potential
energy contributions, see (\ref{intpotrel}).

\section{ Gas of free resonances}\label{Gas}
\subsection{Free relativistic fermion resonances}

Specifics of the choice of the interaction between particles
enters  expression (\ref{energymom}) through a difference $
\epsilon^{\rm int}_{\rm f}-\epsilon^{\rm pot}_{\rm f}$. We
introduce the term "free resonances", describing the case, when
one may neglect the value $ \epsilon^{\rm int}_{\rm
f}-\epsilon^{\rm pot}_{\rm f}$. In this section let us consider
thermodynamics of "free resonances". Thus we will use Eqs.
(\ref{nd}), (\ref{ed}), (\ref{p}), (\ref{st}) disregarding  in
(\ref{ed}), (\ref{p}) $\epsilon^{\rm int}_{\rm f}-\epsilon^{\rm
pot}_{\rm f}$ terms.

 In the limit of vanishing density, dependence of diagrams determining the resonance
width on particle occupations (on $\mu$ and $T$) ceases. Then for
relativistic particles  the
 spectral function depends on the energy and momentum
 through  $s=\om^2 -p^2$ variable, cf. \cite{Leupold}. (For
non-relativistic particles, instead of $s$ we would use $\om
-\om_p^{\rm n.r.}$.) Thus, dealing here with relativistic fermions
we may assume that

\be\label{anz}\widetilde{A}_{\rm f}=\widetilde{A}_{\rm f}(s),
\quad s=\om^2 -p^2 ,\ee where $\widetilde{A}_{\rm f}$ is related
to the spectral function following Eq. (\ref{an}).

In the medium $\widetilde{A}_{\rm f}$ may depend on $\om$ and
$\vec{p}$ separately. Indeed, Green functions enter  diagrams,
that determine the  self-energy of the resonance, together with
particle occupations. Thus, also $\widetilde{A}_{\rm f}$ depends
on $\mu$ and $T$. At a finite density  the full width can be
presented as sum of the vacuum and the medium terms. The latter
term ceases with decrease of the density. Thereby in the low
density limit we can use expression (\ref{anz}).

Now we are able to check that  our expressions for thermodynamic
quantities of "free resonances"  fulfill thermodynamic consistency
conditions (\ref{consist}). Taking derivatives of the pressure
(\ref{p})  we use
 partial integrations and relations
 \be\label{help}
\frac{\partial \widetilde{A}_{\rm f} }{\partial \om^2
}=-\frac{\partial \widetilde{A}_{\rm f}}{\partial \vec{p}^{\,2}} ,
\quad \frac{\partial f_{\rm f}}{\partial T}=-\frac{(\om -\mu )}{T}
\frac{\partial f_{\rm f}}{\partial \om}, \quad \frac{\partial
f_{\rm f}}{\partial \mu}=-\frac{\partial f_{\rm f}}{\partial
\om}.
 \ee
Doing partial integrations we drop "surface terms" that requires a
smooth switching of the resonance spectral function at the
threshold (due to switching of the width) and at infinity:
 \be\label{thres} A_{\rm f}(s\rightarrow s_{\rm th})\rightarrow 0,
\quad A_{\rm f}(s\rightarrow 0)\rightarrow 0, \quad A_{\rm
f}(s\rightarrow \infty)\rightarrow 0.
 \ee
Thus we arrive at expressions (\ref{nd}) and (\ref{st}).

Also, a convenient expression for the entropy density can be then
recovered with the help of expressions (\ref{consist}),
(\ref{presmu}), and the relation
 \be\label{repl}\frac{\partial f_{\rm f}}{\partial
T}=-\frac{\partial \sigma_{\rm f}}{\partial \om}.
 \ee
We again use partial integrations and relations (\ref{help}).
Finally we arrive at expression
 \be\label{end} S_{\rm f}=
N_{\rm f}\int_{0}^{\infty}\frac{4\pi p^2 d
p}{(2\pi)^3}\int_{0}^{\infty}\frac{ d \om}{2\pi} A_{\rm f}^{(+)}
\sigma_{\rm f}^{(+)}+(\mu_{\rm f}\rightarrow -\mu_{\rm f }).
 \ee
Here
 \be\label{sigmr} \sigma_{\rm f}^{(+)}=-(1-f_{\rm f}^{(+)})\ln
(1-f_{\rm f}^{(+)}) -f_{\rm f}^{(+)}\ln f_{\rm f}^{(+)}
 \ee
 is the
entropy distribution function. Doing similar manipulations we
derive expression (\ref{end}) from  (\ref{st}), that again proves
thermodynamical consistency of our derivations.

 Note that we used only assumption
$\epsilon^{\rm int}_{\rm f}-\epsilon^{\rm pot}_{\rm f}=0$, an
ansatz (\ref{an}), (\ref{anz}) for the spectral function and
conditions (\ref{thres}) to check that  thermodynamical quantities
that we have derived satisfy thermodynamical consistency
conditions. On the other hand, Refs. \cite{IKV1,IKV2} have shown
that,  even with the full interaction  included, the baryon
density is presented as the sum of partial Noether contributions.
Since we proved thermodynamic consistency of our "free resonance"
model and recovered Noether quantities, we could hope that at
least in the low density limit our ansatz (\ref{an}), (\ref{anz})
and assumption,  that terms $\epsilon^{\rm int}_{\rm
f}-\epsilon^{\rm pot}_{\rm f}$ being present in the pressure do
not contribute to the particle  density, are indeed justified, i.e
$\delta n= -\frac{\partial (\epsilon^{\rm int}_{\rm
f}-\epsilon^{\rm pot}_{\rm f})}{\partial \mu}= 0$. However with
the help of partial integrations  one can show that $\delta n=
-\frac{\partial (\epsilon^{\rm int}_{\rm f}-\epsilon^{\rm
pot}_{\rm f})}{\partial \mu}\neq 0$ and thus there appears an
extra contribution to the Noether density. Also the entropy
density acquires extra term $\delta S=-\frac{\partial
(\epsilon^{\rm int}_{\rm f}-\epsilon^{\rm pot}_{\rm f})}{\partial
T}$. On the other hand, since the entropy density presented in the
form   (\ref{st}) should not depend on the interaction terms, we
may conclude that $\delta n$ and $\delta S$ should be proportional
to the density in a higher power than the Noether terms  and those
interaction terms can be indeed neglected in the virial limit.
Below, on explicit examples we show that $\delta n \rightarrow 0$
and $\delta S\rightarrow 0$ in the virial limit.

Anyhow, the model of "free resonances" continues to be
thermodynamically consistent even at a higher density, where its
results, of course, should deviate from exact solutions.

To do the problem  tractable, instead of solving a complete set of
Dyson equations, we may select a simplified phenomenological
expression for $A_{\rm f}$ (compare with \cite{Leupold,KTV2}),
e.g.,
 \be\label{fenomen} A_{\rm f} =\frac{2\xi \om [2\Gamma_{\rm
f} (s)+\delta] }{(s -m^{*2}_{\rm f})^2 +[\Gamma_{\rm r} (s)+\delta
]^2 }, \quad \xi =const,\quad s=\om^2 -p^2
>s_{\rm th}>0,
 \ee
for $\delta \rightarrow 0$, with a simple $s$-dependence of the
width, e.g.,
 \be\label{widthf}\Gamma_{\rm f} (s)=\Gamma_0 m_{\rm
r}^{1-2\alpha} F(s) (s-s_{\rm th})^{\alpha}\theta (s-s_{\rm
th}),\quad \Gamma_0 =const.
 \ee
 Here  $\alpha =l+1/2$, i.e., $\alpha =1/2$ for the $s$-wave resonance and
 $\alpha$
=3/2 for the $p$-wave resonance. The  width should vanish at the
threshold,
 i.e. $\alpha >0$.
An extra form-factor, $F(s)$, is introduced to correct the
high-energy behavior of the width. One can take $F^{-1}=
1+[(s-s_{\rm th})/\Lambda^2]^\beta$ with $s_0$ and $\beta$ being
constants. Appropriate values for the cut-off factor $\Lambda$ and
the power $\beta$ are: $\Lambda\sim (0.7\div 1)$ GeV and $\beta
>\alpha +1/2$. The latter inequality  provides a decrease of the width at large
energies, far off the resonance peak. If a more detailed
description is required, one can use a more involved
phenomenological expression for the width fitting free parameters
from comparison of the resonance shape with experimental data.

 The energy dependence of the width may cause
a problem. With a simple ansatz (\ref{widthf}) for the behavior of
$\Gamma_{\rm f} (s)$ we get a complicated  $m^{*2}_{\rm f} (s)$
dependence of the effective resonance mass, as it follows from the
Kramers-Kronig relation. However, using that $m^{*2}_{\rm f}\neq
0$ at the threshold and that  $m^{*2}_{\rm f} (s)$ is a smooth
function of $s$, we may ignore mentioned complexity and put for
simplicity $m^{*2}_{\rm f}\simeq const.$  Factor $\xi$ is
introduced to fulfill the sum-role
 \be\label{sr}
\int_{0}^{\infty}\frac{ \rmd s }{4\pi} \widetilde{A}_{\rm f} =1
,
 \ee
that yields $\xi \simeq 1+O(\Gamma_0 /m^{*}_{\rm f} )$ (for
$m^{*}_{\rm f}\gg \Gamma_0 $, $m^{*}_{\rm f}>s_{\rm th}$).
Closeness of $\xi$ to unity (for $ \Gamma_0 /m^{*}_{\rm f}\ll 1$)
shows that exact form of the spectral function (\ref{A-G})   is
corrected by $\xi$ only slightly.

The baryon number density, the energy density, the pressure and
the entropy density
   are determined by  expressions  (\ref{nd}), (\ref{ed}), (\ref{p}),
 (\ref{end}) (where for "free resonances"  $\epsilon_{\rm f}^{\rm int}-\epsilon_{\rm f}^{\rm pot}=0$).

\subsection{Deficiency of approximation of constant particle
width}\label{Deficiency}

 Note that some thermal models have used
the same expressions for $n$ and $P$ as we derived for "free
resonances". However, these expressions were not derived in those
models but introduced using intuitive arguments like a smearing of
the $\delta$-function spectral density. Then
 for simplicity one often applies approximation of constant particle width.
Here we would like to pay attention to the fact that  with
assumption of constant width ($\Gamma_{\rm f} =\Gamma_0 =const.$
instead of (\ref{widthf})) conditions (\ref{thres}) are not
anymore fulfilled and the model would suffer of thermodynamical
inconsistency, if one used mentioned expressions for
thermodynamical quantities. E.g., if one found the particle
density and the entropy density with the help of thermodynamical
consistency conditions (\ref{consist}) using Eq. (\ref{p}) for the
pressure at $\epsilon_{\rm f}^{\rm int}-\epsilon_{\rm f}^{\rm
pot}=0$ and at assumption of the constant width, there would
appear  extra "surface integral" contributions
 \be
&&\Delta n_{\rm f} = N_{\rm f}\int_{0}^{\infty}\frac{4\pi p^4 d
p}{(2\pi)^4}[\widetilde{A}_{\rm f}^{(+)} f_{\rm
f}^{(+)}]|_{\sqrt{s_{\rm th}+p^2}}  \nonumber\\ &&-
\frac{N_{\rm f}}{3}\int_{0}^{\infty}\frac{ d
\om}{2\pi^2}\widetilde{A}_{\rm f}^{(+)}(s=0) f_{\rm f}^{(+)}\om^4
- (\mu_{\rm f }\rightarrow -\mu_{\rm f }),
 \ee
 \be &&\Delta S_{\rm th} =-
\frac{N_{\rm f}}{3}\int_{0}^{\infty}\frac{4\pi p^2 d
p}{(2\pi)^4}[\widetilde{A}_{\rm f}^{(+)} \sigma_{\rm
f}^{(+)}]|_{\om_{\rm th}}p^2 \nonumber\\ &&- \frac{N_{\rm
f}}{3}\int_{0}^{\infty}\frac{ d \om}{2\pi^2}\widetilde{A}_{\rm
f}^{(+)}(s=0) \sigma_{\rm f}^{(+)}\om^4 + (\mu_{\rm f}\rightarrow
-\mu_{\rm f }),
 \ee
compared to Noether values (\ref{nd}) and (\ref{end}).  These
extra terms would vanish, if conditions (\ref{thres}) were
fulfilled.

On the other hand, instead of  derivation of expression for the
particle density from expression for the pressure (or $\Omega$),
one could start with exact Noether expression for the particle
density. Then  the pressure and other thermodynamic
characteristics would acquire extra "surface integral"
corrections, if one assumed constant widths.

\subsection{Free relativistic boson  resonances}

 Using ansatz $A=A(s)$ and doing the same replacements of variables
(\ref{help}), (\ref{repl}), as before, we easily check
thermodynamical consistency of  expressions for  thermodynamical
quantities: the energy density, the pressure and the entropy
density (in a convenient form)
 \be
E_{\rm b}=N_{\rm b}\int_{0}^{\infty}\frac{4\pi p^2 d
p}{(2\pi)^3}\int_{s_{\rm th}}^{\infty}\frac{ d s}{2\pi}\om A_{\rm
b}^{(+)} f_{\rm b }^{(+)}+(\mu_{\rm b}\rightarrow -\mu_{\rm b}),
 \ee
 \be
P_{\rm b}=N_{\rm b}\int_{0}^{\infty}\frac{4\pi p^2 d
p}{(2\pi)^3}\int_{{s_{\rm th}}}^{\infty}\frac{ d
s}{2\pi}\frac{{p}^2}{3\om}A_{\rm b}^{(+)} f_{\rm b}^{(+)}
+(\mu_{\rm b }\rightarrow -\mu_{\rm b}),
 \ee
 \be
S_{\rm b}=N_{\rm b}\int_{0}^{\infty}\frac{4\pi p^2 d
p}{(2\pi)^3}\int_{{s_{\rm th}}}^{\infty}\frac{ d s}{2\pi} A_{\rm
b}^{(+)} \sigma_{\rm b}^{(+)}+(\mu_{\rm b}\rightarrow -\mu_{\rm b
}),
 \ee
now with
 \be\label{sigmrb} \sigma_{\rm b}^{(+)}=(1+f_{\rm
b}^{(+)})\ln (1+f_{\rm b}^{(+)}) -f_{\rm b}^{(+)}\ln f_{\rm
b}^{(+)}.
 \ee
$N_{\rm b}=1$ for spin-less bosons.

 For practical
calculations we may use \cite{KTV2},
 \be\label{fenomen1} A_{\rm
b}^{(\pm)} =\frac{\xi
 [\Gamma_{\rm b} (s)+\delta] }{(s -m^{*2}_{\rm b})^2 +[\Gamma_{\rm b}
(s)+\delta ]^2 /4 }, \quad \xi =const,
 \ee
 \be\label{widthb}\Gamma_{\rm b} (s)=2\Gamma_0 m_{\rm
r}^{1-2\alpha}F(s) (s-s_{\rm th})^{\alpha}\theta (s-s_{\rm
th}),\quad \Gamma_0 =const.
 \ee

Thus, taking into account additional pre-factor $v^0 =2\om$, that
appears in the bosonic case, we see that  the bosonic and
fermionic spectral functions (\ref{fenomen1}), (\ref{fenomen}) are
similar.

\section{Examples of the description of some systems of
interacting  particles with widths}\label{System}

Above we used ansatze for spectral functions.  Also we neglected
$\epsilon_{\rm int}-\epsilon_{\rm pot}$ term. These assumptions
are not fulfilled in case of dense systems.
 Therefore it is worthwhile
to obtain exact relations for thermodynamic quantities  not doing
any assumptions. Below consider the following examples of
interacting systems: of one particle species with non-relativistic
paired interaction; of single relativistic boson species with
$\phi^4$ self-interaction; of two particle species with two
single-flavor fermion interacting via one-boson exchange, as
$NN\sigma$; and of three species with two-different flavor
fermions interacting via one boson exchange, as $\Delta N\pi$.

Note that  in most practically interesting situations a broad
resonance appears, as a consequence of the interaction between
other particle species. Those (other) particle species in many
cases acquire much smaller widths than the given broad resonance
and thereby they can be treated within the quasiparticle
approximation. As example, we may refer  to the $\Delta N\pi$
system. In the latter case the $\Delta$ is a broad resonance (the
width in vacuum reaches 115 MeV) but nucleons and pions have much
smaller widths, at least at not too high temperature. Therefore in
many situations nucleons and pions can be  considered within the
quasiparticle approximation. These observations help us to
construct a simplified treatment of the problem.

{\subsection{Non-relativistic particles interacting via paired
potential}}

 Let us start with discussion of example of the system
of non-relativistic  particles of single species interacting via
paired potential ($\alpha =4$ in (\ref{intpotrel})). Using
(\ref{enrel}), (\ref{pnrel}) and (\ref{pair}) without doing any
approximations we find
 \be E_{\rm n.r.}=
N_{\rm n.r.}\int_{0}^{\infty}\frac{4\pi p^2 d
p}{(2\pi)^3}\int_{\om_{\rm th.}}^{\infty}\frac{ d \om}{2\pi}
\left[\om -\frac{(\om -\om_p^{\rm n.r.} )}{2}\right] A_{\rm n.r.}
f,
 \ee
 \be \label{pairedp}P_{\rm n.r.}=
 N_{\rm n.r.}\int_{0}^{\infty}\frac{4\pi p^2 d
p}{(2\pi)^3}\int_{\om_{\rm th.}}^{\infty}\frac{ d \om}{2\pi}
\left[\frac{2\om_p }{3}+\frac{(\om -\om_p^{\rm n.r.} )}{2}
\right]A_{\rm n.r} f.
 \ee
Thermodynamical relations (\ref{consist}) should be fulfilled with
exact spectral function, since we did not do yet any
approximations. However it is not so easy to satisfy these
relations, if one uses a simplified phenomenological ansatze
spectral functions instead of the exact ones. E.g., conditions
(\ref{consist}) would be violated, if one used  the above
presented ansatze spectral functions applying Eq. (\ref{pairedp})
for the case of a dense system.

Within the quasiparticle approximation at low density one has
$A_{\rm n.r.}\rightarrow (2\pi)\delta (\om -\om_p^{\rm n.r.} )$
and the interaction term $\epsilon_{\rm int}-\epsilon_{\rm pot}$
ceases. By this we explicitly demonstrate that the "free resonance
term" yields a dominant contribution in the low density limit, at
least, if the particle width is sufficiently small. Thereby we
arrive at {\em{a consistent treatment of the problem for a low
density system in the case, when the width is finite but small.}}

{\subsection{Bosonic $\phi^4$ theory}

Using  $\lambda\phi^4$ coupling we may estimate interactions in
the pion gas. To be specific consider here spin-less neutral
bosons ($\mu_{\rm b} =0$). As in case of the paired potential,
here $\alpha =4$ in (\ref{intpotrel}). With the help of Eqs.
(\ref{Ebo}), (\ref{Pbo}), and (\ref{f-b}) we find
 \be E_{\rm
b}=\int_{0}^{\infty}\frac{4\pi p^2 d
p}{(2\pi)^3}\int_{0}^{\infty}\frac{ d \om}{2\pi} \left[2\om^2 -
\frac{(\om^2 -\om^2_p)}{2}\right] A_{\rm b} f_{\rm b},
 \ee
 \be P_{\rm b}=\int_{0}^{\infty}\frac{4\pi p^2 d
p}{(2\pi)^3}\int_{0}^{\infty}\frac{ d \om}{2\pi}
\left[\frac{2p^2}{3}+\frac{(\om^2 -\om_p^2  )}{2}\right] A_{\rm b}
f_{\rm b}.
 \ee
As in case with paired potential, if we used a simplified
phenomenological ansatze spectral functions instead of the exact
ones, we would retain with the same problem with fulfillment of
the consistency conditions for a dense system. In the low density
limit we again deal with free resonances, at least provided
particle width is finite but rather small.

{\subsection{Two single-flavor fermions interacting via one-boson
exchange}} In this case for relativistic fermions and neutral
bosons
 \be &&TS_{\rm f}+TS_{\rm b}=
 N_{\rm f} \int_{0}^{\infty}\frac{4\pi p^2_{\rm f} d p_{\rm
f}}{(2\pi)^3}\int_{0}^{\infty}\frac{ d \om_{\rm f}}{2\pi}
\left(\om_{\rm f} +\frac{p^2_{\rm f}}{3\om_{\rm f}}\right)
 A_{\rm f}^{(+)} f_{\rm f}^{(+)}-\mu_{\rm f}^{(+)}n_{\rm f}^{(+)}\nonumber\\
&+&(\mu^{(+)}_{\rm f}\rightarrow -\mu^{(-)}_{\rm f})\nonumber\\
&+&N_{\rm b}\int_{0}^{\infty}\frac{4\pi p^2_{\rm b} d p_{\rm
b}}{(2\pi)^3}\int_{0}^{\infty}\frac{ d \om_{\rm b}}{2\pi}
\left(2\om_{\rm b}^2 +\frac{2p^2_{\rm b}}{3}\right) A_{\rm
b}^{(+)} f_{\rm b}^{(+)}. \nonumber
 \ee
 Here to simplify expression
we consider the case of $\mu_{\rm b}=0$. For spin-less bosons
$N_{\rm b}=1$.

 Using that following
(\ref{twoFoneB}) $\epsilon_{\rm int}-\epsilon_{\rm pot}=
-\epsilon_{\rm pot}^{\rm f} /2$, and Eq. (\ref{pairR}) for
relativistic fermions, we find
 \be\label{fE} &&E_{\rm f}+E_{\rm
b}= N_{\rm f}\int_{0}^{\infty}\frac{4\pi p^2_{\rm f} d p_{\rm
f}}{(2\pi)^3}\int_{0}^{\infty}\frac{ d \om_{\rm f}}{2\pi}
\left[\om_{\rm f} -\frac{(\om_{\rm f} -\om_p^{\rm f} )}{2}\right]
 A_{\rm f}^{(+)} f_{\rm f}^{(+)}\nonumber\\
 &+&(\mu^{(+)}_{\rm f}\rightarrow -\mu^{(-)}_{\rm f})+N_{\rm
b}\int_{0}^{\infty}\frac{4\pi p^2_{\rm b} d p_{\rm
b}}{(2\pi)^3}\int_{0}^{\infty}\frac{ d \om_{\rm b}}{2\pi}
2\om_{\rm b}^2  A_{\rm b}^{(+)} f_{\rm b}^{(+)}
  ,
   \ee
 \be\label{fP} &&P_{\rm f}+P_{\rm b}=
N_{\rm f}\int_{0}^{\infty}\frac{4\pi p^2_{\rm f} d p_{\rm
f}}{(2\pi)^3}\int_{0}^{\infty}\frac{ d \om_{\rm f}}{2\pi}
\left[\frac{p^2_{\rm f}}{3\om_{\rm f}}+\frac{(\om_{\rm f}
-\om_p^{\rm f} )}{2}\right]
 A_{\rm f}^{(+)} f_{\rm
f}^{(+)}\nonumber\\  &+&(\mu^{(+)}_{\rm f}\rightarrow
-\mu^{(-)}_{\rm f})+N_{\rm b}\int_{0}^{\infty}\frac{4\pi p^2_{\rm
b} d p_{\rm b}}{(2\pi)^3}\int_{0}^{\infty}\frac{ d \om_{\rm
b}}{2\pi} \frac{2p^2_{\rm b}}{3} A_{\rm b}^{(+)} f_{\rm b}^{(+)}.
 \ee
As we see, the interaction affects the fermion contribution,
whereas the boson is described, as the "free resonance" or "free
blurred particle". For fermions expressions look similar to the
case of the system with paired interaction.

These expressions are very convenient in case when {\em{ fermions
are good quasiparticles but bosons might be broad resonances or
blurred particles.}} In the low density limit, using the
quasiparticle spectral function for fermions we see that
{\em{interaction contributions cease}} and we retain with a broad
boson resonance and the fermion quasiparticle, which interaction
with each other is suppressed.

 On the other hand,
 we may correct  boson contributions by the interaction.
 Using that following
(\ref{twoFoneB}) $\epsilon_{\rm int}=\epsilon_{\rm pot}^{\rm f}$,
i.e. $\epsilon_{\rm int}-\epsilon_{\rm pot}= -\epsilon_{\rm
pot}^{\rm b}$, and Eq. (\ref{f-b})  we find
 \be\label{ebf}
&&E_{\rm f}+E_{\rm b}= N_{\rm f}\int_{0}^{\infty}\frac{4\pi
p^2_{\rm f} d p_{\rm f}}{(2\pi)^3}\int_{0}^{\infty}\frac{ d
\om_{\rm f}}{2\pi} \om_{\rm f}  A_{\rm f}^{(+)} f_{\rm
f}^{(+)}+(\mu^{(+)}_{\rm f}\rightarrow -\mu^{(-)}_{\rm
f})\nonumber\\ &+& N_{\rm b}\int_{0}^{\infty}\frac{4\pi p^2_{\rm
b} d p_{\rm b}}{(2\pi)^3}\int_{0}^{\infty}\frac{ d \om_{\rm
b}}{2\pi} \left[ 2\om_{\rm b}^2 -(\om^2_{\rm b}-(\om_{p}^{\rm
b})^2 )\right] A_{\rm b}^{(+)} f_{\rm b}^{(+)} ,
 \ee
 \be\label{pbf} &&P_{\rm f}+P_{\rm b}=
N_{\rm f}\int_{0}^{\infty}\frac{4\pi p^2_{\rm f} d p_{\rm
f}}{(2\pi)^3}\int_{0}^{\infty}\frac{ d \om_{\rm f}}{\pi}
\frac{p^2_{\rm f}}{3\om_{\rm f}} A_{\rm f}^{(+)} f_{\rm
f}^{(+)}+(\mu^{(+)}_{\rm f}\rightarrow -\mu^{(-)}_{\rm
f})\nonumber\\ &+&N_{\rm b}\int_{0}^{\infty}\frac{4\pi p^2_{\rm b}
d p_{\rm b}}{(2\pi)^3}\int_{0}^{\infty}\frac{ d \om_{\rm b}}{2\pi}
\left(\frac{2p^2_{\rm b}}{3}+\om^2_{\rm b}-(\om_{p}^{\rm
b})^2\right) A_{\rm b}^{(+)} f_{\rm b}^{(+)}.
 \ee
Here interaction affects the boson contribution, whereas the
fermion is described, as the "free resonance" or "free blurred
particle". For the boson sub-system expressions look the same, as
for the $\phi^4$ theory.

Again, in the low density limit  the interaction terms cease in
(\ref{ebf}), (\ref{pbf}) provided bosons are good quasiparticles.
The problem then is reduced to description of a broad boson
resonance and the fermion quasiparticle, not interacting with each
other.

\subsection{Two different-flavor fermions interacting via one-boson
exchange}

 For a theory with two different-flavor fermion --
one boson coupling, using (\ref{twodFoneB}) one gets
 \be\epsilon_{\rm int}-\epsilon_{\rm pot}=-\epsilon_{\rm pot}^{\rm f1}=-\epsilon_{\rm pot}^{\rm f2}
=-\epsilon_{\rm pot}^{\rm b} .
 \ee
Thus compared to  previous example,  two particle species can be
described as "free resonances" or "free blurred particles",
whereas the third particle species contains the interaction term.
In case of the $\Delta -N-\pi$ system at not too high baryon
density and temperature, only $\Delta$ acquires a large width.
Thereby we may treat the $\Delta$, as a broad  "free resonance"
with the interaction affecting either nucleon or  pion
sub-systems. Having smaller widths, both nucleon and pion
sub-systems can be described in the quasiparticle approximation.
Nevertheless, if one wanted to proceed further and incorporate the
nucleon (or the pion) width, one could transport the interaction
to the sub-system of particles with the smallest width, treating
only the latter sub-system within the quasiparticle approximation.
 This way we may  solve a complicated problem, being sure that
thermodynamic consistency conditions are approximately fulfilled.

The following remark is in order. Considering $\pi N\Delta$ system
in the virial limit and assuming that only $g_{\pi N\Delta}$
coupling retains, Ref. \cite{Weinhold} constructed explicit
expressions for thermodynamic potential and the net particle
density, being expressed through the so called $B$-spectral
function of the $\Delta$  that depends on $\re\Sigma_{\Delta}$ and
$\im\Sigma_{\Delta}$ in a more complicated way than the standard
$A$-function. On the other hand, the nucleon and the pion
sub-systems are described there, as  ideal gases of free
particles. In spite of apparent difference with obtained above
expressions, both appoaches coincide with each other. Indeed,
Refs. \cite{IKV1,IKV2} have shown that expression for the total
particle density, as the sum of the Noether densities (expressed
entirely in terms of the $A$-spectral functions) fulfills
thermodynamical consistency condition (first condition
(\ref{consist})). Namely using the latter condition Ref.
\cite{Weinhold} has derived the net particle density presented
there in a different form (using the $B$-function). The
$B$-function is expressed in terms of the phase shifts in the
$S$-matrix formulation of statistical mechanics \cite{DMB}.

\section{Hot matter
consisting of strongly interacting light bosons and heavy fermions
at zero chemical potentials}\label{Quasi}

In sect. \ref{Gas} we  demonstrated how one can approximately
describe dilute heated matter with the help of  simple
parameterizations of spectral functions. Now consider example,
where one can find spectral functions and thermodynamic values of
a system of strongly interacting particles avoiding assumption of
dilute matter.

Let us consider hot system of two particle species. One species is
a light boson and another species is a heavy fermion, both at zero
chemical potentials. Let us treat the system in terms of the
self-consistent $\Phi$ derivable approximation scheme. Then the
$\Phi$ functional of Baym is given by diagrams
 \be\label{phi}
\includegraphics[width=7cm,clip=true]{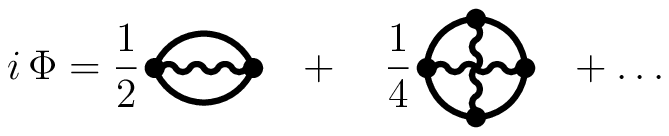}
 \ee
Here both fermion (solid line) and boson (wavy line) Green
functions are full Green functions whereas vertices are bare. Let
us restrict ourselves by consideration of the simplest $\Phi$ (the
first diagram (\ref{phi})). Then the fermion self-energy is as
follows
 \be\label{selfzf}
\includegraphics[width=2cm,clip=true]{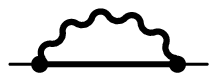}
 \ee
and the boson self-energy reads
 \be\label{selfzb}
\includegraphics[width=2cm,clip=true]{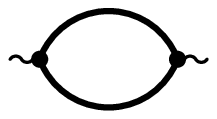}
 \ee
All the multi-particle rescattering processes
 \be\label{ladpr}
\includegraphics[width=7cm,clip=true]{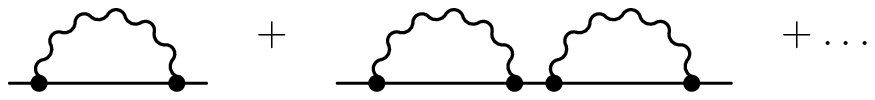}
 \ee
are then included, whereas processes with  crossing of boson lines
(correlation effects) like
 \be\label{cr}
\includegraphics[width=3cm,clip=true]{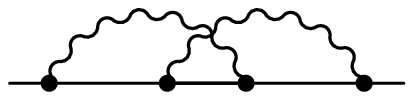}
 \ee
 are disregarded.

 To simplify derivations consider
example of the Yukawa interaction of spin $1/2$ non-relativistic
heavy fermion with a light relativistic scalar boson. The
interaction Lagrangian is given by
 \be\label{intLag} L_{\rm
int}=g\bar{\psi}\phi\psi .
 \ee

Examples of different couplings of relativistic fermions and
bosons, including the Yukawa interaction, can be found in
\cite{V04}. However there we did not  calculate thermodynamic
quantities, that is our aim now.

\subsection{Blurred fermions}

Since $m_{\rm f} \gg m_{\rm b}$, one may expect that in a broad
temperature range, which we will interested in, {\em{boson
occupations are essentially higher than fermion ones}}. Then, at
such temperatures we may retain in (\ref{selfzf}) only terms
proportional to boson occupations. Using this we find \cite{V04}:
 \be\label{sigm-R0-sim} \Sigma_{\rm f}^{R}(p_{\rm f}) &\simeq&\int
g^2 \frac{d^3 p_{\rm b}}{(2\pi)^3} \int_{0}^{\infty} \frac{d
\om_{\rm b}}{2\pi} [G_{\rm f}^{R} (p_{\rm f}+p_{\rm b}) +G_{\rm
f}^{R} (p_{\rm f}-p_{\rm b})] {A}_{\rm b}(p_{\rm b})f_{\rm
b}(\om_{\rm b} ).
 \ee
Let us use {\em the soft thermal loop} (STL) approximation
\cite{Dyug,V04} resulting in dropping a $p_{\rm b}$-dependence of
fermion Green functions in (\ref{sigm-R0-sim}). Then Eq.
(\ref{sigm-R0-sim}) is simplified as
 \be\label{sigmJ}
{\Sigma}_{\rm f}^{R}(p_{\rm f})\simeq {J}\cdot {G}_{\rm
f}^{R}(p_{\rm f}), \quad J=2g^2 \int \frac{d^3 p_{\rm
b}}{(2\pi)^3}\int_{0}^{\infty} \frac{d \om_{\rm b}}{2\pi} A_{\rm
b}(p_{\rm b}) f_{\rm b}(\om_{\rm b} ).
 \ee
As we will see, at $T\gsim m_{\rm b}^{*}(T)$ (further we will
consider namely such temperatures) departure of the fermion energy
from the mass shell $\delta \om_{\rm f} \sim \sqrt{J}$ is much
larger than that for bosons, $\delta \om_{\rm b}\sim \mbox{max}\{
m_{\rm b} -m_{\rm b}^* (T), T\}$, and typical fermion momenta
${p}_{\rm f}\sim \sqrt{2 m_{\rm f}T}$ are much higher than typical
boson momenta ${p}_{\rm b}\sim \mbox{max}\{\sqrt{2 m_{\rm b}^*
(T)T},\,T\}$. Here $m_{\rm b}^* (T)$ is an effective boson mass.
At these conditions the STL approximation should be valid.
Moreover, we assume that $\sqrt{2 m_{\rm f}T}\ll m_{\rm f}$ and
$\sqrt{J}\ll m_{\rm f}$. Then fermions can be treated as
non-relativistic particles. The latter approximation allows us to
avoid a complicated spin algebra.

As follows from (\ref{sigmJ}), the quantity ${J}$ can be expressed
through the tadpole diagram
 \be\label{ladprtad}
\includegraphics[width=2cm,clip=true]{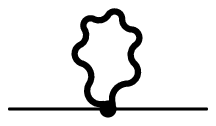}
 \ee
 The latter does not depend on the energy-momentum transfer.
It describes fluctuations of virtual (off-mass shell) bosons. On
the other hand, $J$ can be  interpreted as the density of
quasi-static boson impurities.  Due to multiple repetition of this
diagram in the Dyson series for the fermion, $J$ demonstrates {\em
the intensity of the multiple quasi-elastic scattering} of the
fermion on quasi-static boson impurities.

 Dyson equation for the retarded fermion Green function is greatly
 simplified in the STL approximation:
  \be
 G_{\rm f}^R =G_{0{\rm f}}^R + G_{0{\rm f}}^R J (G_{\rm f}^R)^2 ,
 \ee
with a simple analytical solution
 \be\label{tblrel}
 G_{\rm f}^R =\frac{\om_{\rm f} -\om_p^{\rm f} \pm \sqrt{(\om_{\rm f} -\om_p^{\rm f} )^2
 -4J}}{2J}, \quad \om_p^{\rm f} =m_{\rm f} +\frac{p^2_{\rm f}}{2m_{\rm f}}.
  \ee
 In this problem it is convenient to count  fermion energies from the mass-shell.

  Only
negative sign solution satisfies the retarded property and should
be retained.
  For $(\om_{\rm f} -\om_p^{\rm f}
)^2\gg 4 J$ we recover the quasiparticle (pole-like) solution.
Since then typical energies are $\om_{\rm f} -\om_p^{\rm f} \sim
T$, the quasiparticle approximation  is valid for fermions only
for $J \ll T^2$. Otherwise (for $J\gsim T^2$) fermion Green
function is completely regular. {\em{Fermions become  blurred
particles}}. From (\ref{tblrel}), using relation $(G^R)^{-1}_{\rm
f}=\om_{\rm f} -\om_p^{\rm f}-\mbox{Re}\Sigma^R_{\rm
f}+i\Gamma_{\rm f}/2$, for $4J>(\om_{\rm f} -\om_p^{\rm f} )^2$ we
find
 \be\label{tblrel1} &&\mbox{Re}\Sigma^R_{\rm f}
=\frac{\om_{\rm f} -\om_p^{\rm f}}{2},\quad \Gamma_{\rm f}
=\sqrt{4J-(\om_{\rm f} -\om_p^{\rm f} )^2 }\,\,\theta
\left(4J-(\om_{\rm f} -\om_p^{\rm f} )^2\right),\nonumber\\
&&\quad A_{\rm f}=\frac{\sqrt{4J-(\om_{\rm f} -\om_p^{\rm f} )^2
}}{J}\,\,\theta \left(4J-(\om_{\rm f} -\om_p^{\rm f} )^2\right).
\ee

\subsection{Intensity of multiple scattering}
Now we are able to evaluate the intensity of multiple scattering
$J$. Assume that  neutral spin-less bosons under consideration are
good quasiparticles in the energy-momentum and temperature region
of our interest. Then their spectral function is as follows
$A_{\rm b}= 2\pi\delta \left(\om_{\rm b}^2 -{p}_{\rm
b}^{\,\,2}-m_{\rm b}^2 -\re \Sigma_{\rm b}^{R} (\om_{\rm b} ,
{p}_{\rm b})\right)$, cf. Eq. (\ref{qpbos}),
 and from (\ref{sigmJ}) we find
 \be\label{Jexp} &&J =
\frac{g^2}{2\pi^2}\int_{0}^{\infty}\frac{ {p}_{\rm b}^{\,\,2} d
{p}_{\rm b}} { \left[ m_{\rm b}^{*2}(T)+ \beta{p}_{\rm b}^{\,\,2}
\right]^{1/2}}\frac{1}{\mbox{exp}\left[\left( m_{\rm b}^{*2}(T)+
\beta{p}_{\rm b}^{\,\,2}\right)^{1/2} /T\right]- 1} ,\nonumber
 \ee
where in the second line we adopted  simple form of the
quasiparticle spectrum
  \be\label{branch} \om^2_{\rm b} ({p}_{\rm
b},T) \simeq m_{\rm b}^{*2}(T) +\beta (T){p}_{\rm b}^{\,\,2}
+O({p}_{\rm b}^{\,\,4}).
 \ee
As we will show, in a broad temperature region  of our interest
$\beta$ is close to unity and $m_{\rm b}^{*}$ is close to $m_{\rm
b}$. We present
 \be\label{f0} J =\frac{g^2 T^2}{12\beta^{3/2}}r_s
(z), \quad r_s (z)=\frac{6f_0 (z)}{\pi^2 z^2}, \quad
z=\frac{T}{m_{\rm b}^*}.
 \ee
 Numerical evaluation of the integral
(\ref{f0}) is demonstrated in Fig. \ref{fig:f0.eps}.

 In the
limiting case of a high temperature typical values of momenta are
${p}_{\rm b}\sim T \gg m_b^{*}(T)$ and we obtain
 \be\label{Jexp-lim2} J \simeq \frac{g^2  T^2}{12\beta^{3/2}}
\,,\quad \mbox{for} \quad T\gg m_b^{*}(T) .
 \ee
In the limit $T\ll m_{\rm b}^{*}(T)$, the intensity of multiple
scattering is exponentially suppressed
 \be\label{Jexp-lim1} J
\simeq \frac{g^2 T^{3/2}\sqrt{m_{\rm b}^{*}} } {2^{3/2}
\pi^{3/2}\beta^{3/2}} \mbox{exp}\left( -m_{\rm b}^{*}/T\right)
\,,\quad \mbox{for} \quad T\ll m_{\rm b}^{*}(T).
 \ee

\begin{figure*}
\includegraphics[clip=true,width=8cm]{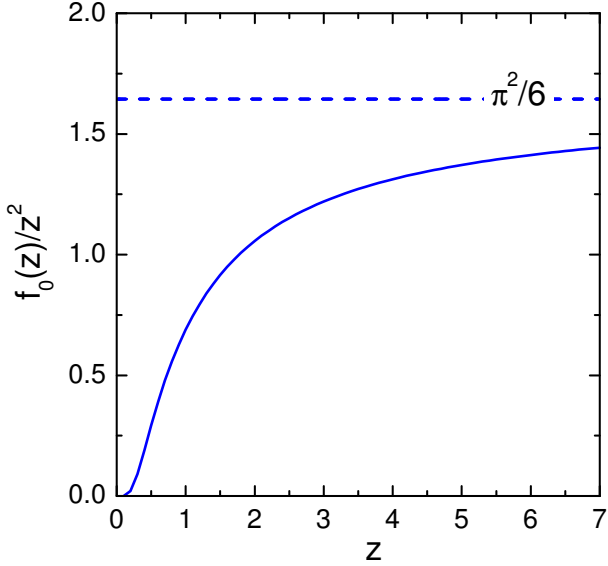}
\caption{$f_0 (z)/z^2$, cf. eq. (\ref{f0}). Dash line demonstrates
asymptotic behavior for $z\gg 1$.} \label{fig:f0.eps}
\end{figure*}
The latter temperature regime is not of our interest here. We are
interested to describe  the temperature region $T\gsim m_{\rm
b}^{*}(T)$, when $J$ is not exponentially suppressed.
 In this case the
quasiparticle approximation (valid for $T^2 \gg 4J$) would work
for fermions only for weak coupling $g \ll \beta^{3/4}\simeq 1$.
Since we are interested in description of strongly interacting
particles, when $g\gsim 1$, we deal with blurred fermions.

The quasiparticle boson density inside the system is given by (we
neglect here a weak boson-energy dependence of the self-energy,
see below):
 \be\label{boso1} n_{\rm b}=
\int_{0}^{\infty}\frac{4\pi p^2 d
p}{(2\pi)^3}\int_{0}^{\infty}\frac{ d \om}{2\pi} 2\om A_{\rm b}
f_{\rm b} \simeq \int_{0}^{\infty} \frac{p^2 dp
}{2\pi^2}\frac{1}{e^{\sqrt{m_{\rm b}^{*\,2} +\beta p^2}/T}-1} ,
 \ee
cf. Eq. (\ref{boso}), whereas for $T\gg m_{\rm b}^{*}$ we find
$n_{\rm b}\simeq T^3 \zeta (3)/(\beta^{3/2}\pi^2)$, $\zeta
(3)\simeq 1.202$.

\subsection{Density of fermion-antifermion pairs}

Consider  system with strong interaction, $g \gsim 1$, and assume
$ m_{\rm b}^* (T)\lsim T\ll 5m_{\rm f}\beta^{3/4}/g $. Then we can
easily check that, on the one hand, conditions of the STL
approximation are fulfilled and, on the other hand, fermions can
be treated as non-relativistic particles. As follows from
(\ref{tblrel}), the fermion spectral function satisfies the full
sum-rule (\ref{sr}). Thus, {\em{although we used approximations,
their consistency is preserved.}}

Following (\ref{nd}) the 3-momentum fermion distribution is as
follows
 \be
{f}_{\rm f}^{(\pm)} (p_{\rm f} )=\int_{0}^{\infty}\frac{d\om_{\rm
f}}{2\pi}A_{\rm f}^{(\pm)}f_{\rm f}(\om_{\rm f}) .
 \ee
Replacing (\ref{tblrel1}) into this expression we find the
3-momentum fermion distribution
 \be\label{hdist} &&{f}_{\rm
f}^{(\pm)} (p_{\rm f} )=\int^{2 \sqrt{J}}_ {-2
\sqrt{J}}\frac{d\xi}{\pi} \frac{\sqrt{4J-\xi^2}}{2J}
\frac{1}{\mbox{exp}[ (\xi +\om_p^{\rm f} )/T]+ 1},
 \ee
where we introduced the variable $\xi =\om_{\rm f} -\om_p^{\rm f}$
and used that $|\xi| <2 \sqrt{J}\ll  m_{\rm f}$. Doing further
replacement $\xi =-2\sqrt{J_s} +Ty$ and using that $m_{\rm f} \gg
T$, we obtain
 \be\label{hdist1} {f}_{\rm f}^{(\pm)} ( p_{\rm f}
)\simeq
 \frac{
T^{3/2}}{\pi J^{3/4}} I \left( \frac{4 \sqrt{J}}{
T}\right)\mbox{exp}\left[-\frac{\om_p^{\rm f} - 2\sqrt{J}
}{T}\right]=: F(T) f_{{\rm Bol}}^{(\pm)}(p_{\rm f} ),
 \ee
 with
  \be
F(T)=\frac{ T^{3/2}}{\pi J^{3/4}} I \left( \frac{4 \sqrt{J}}{
T}\right)\mbox{exp}\left[\frac{2\sqrt{J} }{T}\right],
 \ee
 \be\label{I1} I (x)=\int^{x}_{0} dy e^{-y}\sqrt{y-y^2
x^{-1}},\quad x=4\sqrt{J} /  T, \ee \be\label{I1lim} I (x)&\simeq&
\frac{\pi }{8}x^{3/2}(1-\frac{x}{2}) \,,\quad \mbox{for} \quad
x\ll 1,\\ I (x)&\simeq& \frac{\sqrt{\pi}}{2}
(1-\frac{3}{4x})\,,\quad \mbox{for} \quad x\gg 1. \nonumber
 \ee
As we have mentioned,  for $g \gsim 1$ the condition of the
validity of the quasiparticle approximation for fermions, $x\simeq
4\sqrt{J}/ T \ll 1$, is not fulfilled at temperatures of our
interest.

For  $z\gsim 1$,  and $x\gg 1$ (i.e. for $g\gg 1$), we find
  \be
F(T)\simeq \frac{2^{1/2}3^{3/4}
\beta^{9/8}}{\pi^{1/2}g^{3/2}r_s^{3/4}}\mbox{exp}\left[\frac{gr_s^{1/2}}{3^{1/2}\beta^{3/4}}\right],
 \ee
 and for $x\gg 1$, $z\gg 1$ and $\beta \simeq 1$:
 \be F(T)\rightarrow
 \frac{2^{1/2}3^{3/4}}{\pi^{1/2}g^{3/2}}\mbox{exp}\left[\frac{g}{3^{1/2}}\right]=const
(T) \gg 1  .
 \ee

Integrating (\ref{hdist1}) in momenta we obtain the fermion
(antifermion) density:
 \be\label{ratd} n_{\rm f}^{(\pm)}\simeq
F(T) n_{{\rm Bol}}^{(\pm)}, \quad n_{{\rm Bol}}^{(\pm)}\simeq
N_{\rm f} \left(\frac{m_{\rm f}
T}{2\pi}\right)^{3/2}\mbox{exp}\left[ -\frac{m_{\rm
f}}{T}\right],\quad N_{\rm f}=2.
 \ee

We see that with growing  parameter $\sqrt{J_s}/T$, i.e. with
growing temperature, the density of fermion-antifermion pairs
increases significantly compared to the standard Boltzmann value.

The result (\ref{hdist1}) can be interpreted with the help of the
relevant quantity
 \be\label{effermm} m_{\rm f}^{*} (T) =m_{\rm f}
-2\sqrt{J}, \quad 2\sqrt{J}\ll m_{\rm f},
 \ee
which has the meaning of {\em{the effective fermion  (antifermion)
mass.}} However, contrary to the usually introduced effective
mass, the quantity (\ref{effermm}) enters only the exponent in
expression (\ref{hdist1}). We see that $m_{\rm f}^{* }(T)$
decreases with increase of the intensity of the multiple
scattering $J$, i.e., with increase of the temperature.

Typical temperature, when effective fermion mass decreases
significantly, $\sqrt{J}\sim m_{\rm f}$, is as follows
 \be\label{Tcssn} T\sim T_{\rm bl} \sim \beta^{3/4} g^{-1}m_{\rm
f}r_s^{-1/2} (T_{\rm bl}).
 \ee
For $T\gsim T_{\rm bl}$
 the non-relativistic
approximation for fermions, that we  used, fails. Generalization
to the relativistic case can be found in \cite{V04}. However
exponential increase of the effective fermion mass, that we have
demonstrated, starts already for $T\ll T_{\rm bl}$, in the region
of validity of the non-relativistic approximation. For $g\gg 1$
and $\beta \simeq 1$ we obtain \be T_{\rm bl}\sim m_{\rm f}/g\ll
m_{\rm f}.\ee Thus already for comparatively low temperatures the
fermion vacuum becomes blurred due to strong interaction between
light boson and heavy fermion sub-systems.

\subsection{Boson quasiparticles}
With the help of expression (\ref{tblrel}) we are able to
calculate the boson self-energy (\ref{selfzb}), see \cite{V04}:
 \be
&&\mbox{Re}\Sigma_{\rm b}^R (\om_{\rm b}, p_{\rm b})\simeq -2g^2
\mbox{Tr}\int \frac{d^4 p_{\rm f}}{(2\pi)^4}\left[\mbox{Re}G_{\rm
f}^R (p_{\rm f}+p_{\rm b})+\mbox{Re}G_{\rm f}^R (p_{\rm f}-p_{\rm
b})\right]\nonumber\\ &&\times\mbox{Im}G_{\rm f}^R (p_{\rm
f})f_{\rm f} (\om_{\rm f}) ,
  \ee
 \be
&&\Gamma_{\rm b}^R (\om_{\rm b}, p_{\rm b})\simeq 4g^2
\mbox{Tr}\int \frac{d^4 p_{\rm f}}{(2\pi)^4}\mbox{Im}G_{\rm f}^R
(p_{\rm f}+p_{\rm b})\mbox{Im}G_{\rm f}^R (p_{\rm f})\nonumber\\
&&\times\left[ f_{\rm f} (\om_{\rm f})-f_{\rm f} (\om_{\rm
f}+\om_{\rm b})\right] .
 \ee
Let us present the result in the limit $x\gg 1$ (i.e., $g\gg 1$).
Then we find \cite{V04}:
 \be\label{bm}\mbox{Re}\Sigma_{\rm b}^R
(\om_{\rm b}, p_{\rm b})\simeq -\frac{4g^2 n_{\rm
f}^{(\pm)}}{\sqrt{J}}+\alpha (\om_{\rm b}^2 -\frac{1}{2}p^2_{\rm
b}),\quad \alpha =\frac{4g^2 n_{\rm f}^{(\pm)}}{Jm_{\rm f}},
 \ee
 \be
\Gamma_{\rm b}^R / \mbox{Re}\Sigma_{\rm b}^R =O(T^2 /J).
 \ee
 Using
(\ref{bm}) we recover the effective boson mass \be\label{bosc}
m_{\rm b}^{*\,2} \simeq m_{\rm b}^{2}-\frac{4g^2 n_{\rm
f}^{(\pm)}}{\sqrt{J}}.\ee The value $\alpha$ has extra smallness
for $m_{\rm f}\gg m_{\rm b}$ and dependence on it can be
neglected.

We see that for temperatures $T< T_{\rm bl}$ one has
$|\mbox{Im}\Sigma_{\rm b}^R /\mbox{Re}\Sigma_{\rm b}^R|\ll 1$.
Thus, at these temperatures we may treat bosons in the
quasiparticle approximation. Although the value
$\mbox{Re}\Sigma_{\rm b}^R$ is exponentially
small,\footnote{Following Ref. \cite{V04}, that treated fermions
in relativistic terms, at a somewhat higher temperature, $T>
T_{\rm bl}$, the value $-\re \Sigma_{\rm b}^R$ sharply increases
and the effective boson mass substantially decreases.} we do not
suppress the interaction and do not set $m^*_{\rm b}= m_{\rm b}$
and $\beta \simeq 1$ because, as we show below, the correction due
to the interaction to  boson contributions in some thermodynamic
quantities, e.g.,  pressure, can be larger than the fermion
contributions.

\subsection{Thermodynamic quantities}
Since for temperatures of our interest here,  $T< T_{\rm bl}$,
bosons are good quasiparticles  and fermions are blurred
particles, it is natural to use expressions (\ref{ebf}),
(\ref{pbf}), with interaction affecting boson terms. Then boson
contributions are immediately calculated:
 \be\label{bosqE} E_{\rm
b}^{\rm q.p.}=\int_{0}^{\infty}\frac{4\pi p_{\rm b}^2 d p_{\rm
b}}{(2\pi)^3}f_{\rm b}(\om_{\rm b} (p_{\rm b})) \frac{\om_{\rm
b}^2 (p_{\rm b})-\frac{1}{2}\re \Sigma_{\rm b}^R \left(\om_{\rm b}
(p_{\rm b}), {p}_{\rm b}\right)}{\om_{\rm b} (p_{\rm
b})\left[1-\frac{\partial\re\Sigma_{\rm b}^R}{\partial
\om^2}|_{\om_{\rm b} (p_{\rm b})}\right]},
 \ee
 \be\label{bosqP} P_{\rm b}^{\rm q.p.}=\int_{0}^{\infty}\frac{4\pi
p_{\rm b}^2 d p_{\rm b}}{(2\pi)^3}f_{\rm b}(\om_{\rm b} (p_{\rm
b})) \frac{ \frac{p_{\rm b}^2}{3} +\frac{1}{2}\re \Sigma_{\rm b}^R
(\om_{\rm b} (p_{\rm b}), {p}_{\rm b})}{\om_{\rm b} (p_{\rm
b})\left[1-\frac{\partial\re\Sigma_{\rm b}^R}{\partial
\om^2}|_{\om_{\rm b} (p_{\rm b})}\right]}.
 \ee
Main contributions to the thermodynamic quantities of boson
sub-system are given by non-interacting terms:
 \be\label{bosqEF}
E_{\rm b}^{\rm q.p.}\rightarrow E_{\rm b}^{\rm
free}=\int_{0}^{\infty}\frac{ p_{\rm b}^2 d p_{\rm
b}}{2\pi^2}\om^{\rm b}_pf_{\rm b}(\om^{\rm b}_p) ,\,\,\,\, P_{\rm
b}^{\rm q.p.}\rightarrow P_{\rm b}^{\rm
free}=\int_{0}^{\infty}\frac{ p_{\rm b}^4 d p_{\rm
b}}{6\pi^2\om^{\rm b}_p}f_{\rm b}(\om^{\rm b}_p ) .
 \ee
For $T\gsim m^{*}_{\rm b} (T)$ of our interest we estimate $E_{\rm
b}^{\rm q.p.}\sim 3P_{\rm b}^{\rm q.p.}\sim \pi^2 T^4/30$.

 Let us
consider the  interaction contribution (by interaction term we
mean  the term $\propto \epsilon_{\rm b}^{\rm int}-\epsilon_{\rm
b}^{\rm pot}$):
 \be\label{bosqEcor} \delta E_{\rm b}^{\rm
q.p.}=-\delta P_{\rm b}^{\rm q.p.}\simeq
\int_{0}^{\infty}\frac{4\pi p_{\rm b}^2 d p_{\rm
b}}{(2\pi)^3}f_{\rm b}(\om_{\rm b} (p_{\rm b})) \left[-\frac{\re
\Sigma_{\rm b}^R \left(\om_{\rm b} (p_{\rm b}), {p}_{\rm
b}\right)}{2\om_{\rm b} (p_{\rm b})}\right]. \ee
 We neglected the
value $\frac{\partial\re\Sigma_{\rm b}^R}{\partial
\om^2}|_{\om_{\rm b} (p_{\rm b})}=\alpha \ll 1$ for $g\gsim 1$ and
for temperatures of our interest. Rough estimation yields:
 \be
\delta E_{\rm b}^{\rm q.p.}\sim  -\delta P_{\rm b}^{\rm q.p.}\sim
\frac{g^2 n_{\rm f}^{(\pm)} T^2}{\sqrt{J}}. \ee
 Simple result can
be obtained for $T\gg m^{*}_{\rm b} (T)$:
 \be\label{bosint} \delta
E_{\rm b}^{\rm q.p.}= -\delta P_{\rm b}^{\rm q.p.}\simeq \frac{g^2
n_{\rm f}^{(\pm)} T^2}{6\sqrt{J}}.
 \ee
Now we need to  add to boson quantities (\ref{bosqE}),
(\ref{bosqP})
 the  contribution
of quasi-free  fermion blurs. Integrations in $\om_{\rm f}$ are
done similar to that in (\ref{hdist}). We use that $|\xi| <2
\sqrt{J}\ll  m_{\rm f}$ and thus integrating (\ref{ebf}),
(\ref{pbf}) in $\xi$ we may put $\om_{\rm f}\simeq \om_p^{\rm f}$
 in these expressions in the pre-factor of  the spectral function.
 Therefore,
 \be\label{ofBe}
 E_{\rm f}^{(\pm)}&\simeq&  F(T) E_{{\rm
Bol}}^{(\pm)}, \\
 E_{{\rm Bol}}^{(\pm)}&=&N_{\rm f}
\int_{0}^{\infty}\frac{4\pi p^2_{\rm f} d p_{\rm f}}{(2\pi)^3}
\om_p^{\rm f} f_{{\rm Bol}}^{(\pm)}(p_{\rm f} )\simeq m_{\rm
f}^{*} n_{{\rm Bol}}^{(\pm)}\simeq m_{\rm f} n_{{\rm Bol}}^{(\pm)}
(1-2\sqrt{J}/ m_{\rm f}), \nonumber
 \ee
 \be\label{ofBp} P_{\rm f}^{(\pm)}\simeq  F(T) P_{{\rm
Bol}}^{(\pm)}, \quad P_{{\rm Bol}}^{(\pm)}=N_{\rm f}
\int_{0}^{\infty}\frac{4\pi p^2_{\rm f} d p_{\rm f}}{(2\pi)^3}
\frac{{p}_{\rm f}^2}{3\om_p^{\rm f} } f_{{\rm Bol}}^{(\pm)}(p_{\rm
f} ) \simeq  T n_{{\rm Bol}}^{(\pm)}. \ee
 Thus for the fermion
sub-system we obtained  relations between thermodynamic
quantities, similar to those  for the Boltzmann gas, however with
a common large pre-factor $F(T)$. This pre-factor demonstrates
significantly increased number of produced fermion-antifermion
pairs due to strong interaction between fermion-boson species.
Thus fermion sub-system represents  a gas of blurred fermions.
Additionally we kept a higher order term (for $T\ll T_{\rm bl}$)
in expression for the energy (which is larger than the main term
in the pressure for $\sqrt{J}\gg T$) in order further to fulfill
thermodynamic consistency conditions.

Note that in spite of extra $T$ dependence of the factor $F(T)$ in
(\ref{ofBe}), (\ref{ofBp}) compared to Boltzmann expressions,
 thermodynamic  consistency conditions remain approximately
 fulfilled. Indeed, differentiating $P$  in $T$ in order to check
 validity of the consistency condition for the entropy, see second Eq.
  (\ref{consist}), we may omit terms
 $\propto \sqrt{J}$ compared to $\om_p\simeq m_{\rm f}$. It means
 that with this accuracy we should not differentiate spectral
 function, but only the Bolzmann distribution. In the limit $T\gg
 m^{*}_{\rm b}(T)$, factor $F(T)\rightarrow const (T)$ and it does
 not contribute to the temperature derivative.

On the other hand, we could transport the interaction contribution
to the sub-system of the blurred fermions considering bosons, as
the gas of free quasiparticles. From (\ref{fE}), (\ref{fP}) we
find the interaction term ($\propto \epsilon_{\rm f}^{\rm
int}-\epsilon_{\rm f}^{\rm pot}$):
 \be\label{fEcor} &&\delta
E_{\rm f}=-\delta P_{\rm f}= 2 N_{\rm
f}\int_{0}^{\infty}\frac{4\pi p^2_{\rm f} d p_{\rm
f}}{(2\pi)^3}\int_{0}^{\infty}\frac{ d \om_{\rm f}}{2\pi}
\left[-\frac{(\om_{\rm f} -\om_p^{\rm f} )}{2}\right]
 A_{\rm f}^{(+)} f_{\rm f}^{(+)}  .
  \ee
  Integration is
performed, as in (\ref{hdist}). Then we obtain
 \be \delta E_{\rm
f}=-\delta P_{\rm f}\simeq 2\sqrt{J}n_{\rm f}^{(\pm)},
 \ee
that  for $T\gg m_{\rm b}^*$ coincides with (\ref{bosint}). Thus
we explicitly  demonstrated consistency of our derivations.

For $T\gg m^{*}_{\rm b}(T)$ the "free quasiparticle" contributions
to the boson pressure and the energy are easily calculated
 \be\label{bosqEcorlarge}  &&P_{\rm b}^{\rm free\,\,q.p.}\simeq
\int_{0}^{\infty}\frac{4\pi p_{\rm b}^2 d p_{\rm
b}}{(2\pi)^3}f_{\rm b}(\om_{\rm b} (p_{\rm b})) \frac{p_{\rm
b}^2}{3\om_{\rm b} (p_{\rm b})}\nonumber\\ &&\simeq \frac{\pi^2
T^4}{90}-\frac{m_{\rm b}^{*\,2} T^2}{24} =\frac{\pi^2
T^4}{90}-\frac{m_{\rm b}^2 T^2}{24}+\frac{g^2 n_{\rm f}^{(\pm)}
T^2}{6\sqrt{J}},
 \ee
 \be\label{bosqEcorlargeE}  &&E_{\rm b}^{\rm free\,\,q.p.}\simeq
\int_{0}^{\infty}\frac{4\pi p_{\rm b}^2 d p_{\rm
b}}{(2\pi)^3}\om_{\rm b} (p_{\rm b})f_{\rm b}(\om_{\rm b} (p_{\rm
b})) \nonumber\\ &&\simeq \frac{\pi^2 T^4}{30}-\frac{m_{\rm
b}^{*\,2} T^2}{24}=\frac{\pi^2 T^4}{30}-\frac{m_{\rm b}^2
T^2}{24}+\frac{g^2 n_{\rm f}^{(\pm)} T^2}{6\sqrt{J}}.
 \ee
The total pressure, energy density and the entropy calculated
following Eq. (\ref{presmu})  become
 \be \label{prtot} P_{\rm
tot}\simeq\frac{\pi^2 T^4}{90}-\frac{m_{\rm b}^2 T^2}{24}+2Tn_{\rm
f}^{(\pm)} ,
 \ee
 \be \label{ertot} E_{\rm tot}\simeq\frac{\pi^2
T^4}{30}-\frac{m_{\rm b}^2 T^2}{24}+2m_{\rm f}n_{\rm f}^{(\pm)}
,
 \ee
 \be \label{stot} TS_{\rm tot}= E_{\rm tot}+P_{\rm
tot}\simeq\frac{2\pi^2 T^4}{45}-\frac{m_{\rm b}^2 T^2}{12}+2m_{\rm
f}n_{\rm f}^{(\pm)} .
 \ee
We see that the interaction term $\delta P_{\rm f}=\delta P_{\rm
b}^{\rm q.p.}$ is compensated in the total pressure by the
corresponding boson "free quasiparticle" contribution. On the
other hand, the interaction term in the energy density doubles the
corresponding boson "free quasiparticle" contribution. But this
resulting contribution is compensated by the fermion quasiparticle
term that distinguishes $m_{\rm f}$ from $m_{\rm f}^{*}$. Thus
{\em{the total pressure, the energy and the entropy are
approximately given by the sum of contributions of "free" bosons
and "free" fermion-antifermion blurs.}} With these expressions
thermodynamic consistency conditions  (\ref{consist}) are
fulfilled.

Finally we have arrived at the following picture. A hot system of
strongly interacting ($g\gg 1$) light bosons and heavy fermions
 with zero chemical
potentials at temperatures $T_{\rm bl}>T\gsim m_{\rm b}^* (T)$
represents a   gas mixture  of  boson quasiparticles and blurred
fermions. Blurred heavy fermions undergo rapid ($p_{\rm f}\sim
\sqrt{m_{\rm f}T} \gg p_{\rm b}\sim T$) Brownian motion in the
boson quasiparticle gas. The density of blurred fermions is
dramatically increased at $T\sim T_{\rm bl}$ compared to the
standard Bolzmann value. Thermodynamical quantities are such as
for the quasi-ideal gas mixture of quasi-free fermion blurs and
quasi-free bosons.

\subsection{Boson mean field solution}

The model with the interaction Lagrangian (\ref{intLag}) allows
for the mean field solution for the boson field, cf. Walecka
model. To get this solution one should replace $\phi \rightarrow
\phi_{\rm cl}+\phi$. The classical field can be incorporated into
 expression for the pressure with the help of the replacement  $m_{\rm
f}\rightarrow m_{\rm f}^{\phi}= m_{\rm f}-g\phi_{\rm cl}$ and by
addition of the term $\delta P_{\rm cl}=-\frac{1}{2} m_{\rm
b}^{*\,2}\phi_{\rm cl}^2$. Variation of the pressure in the
classical field produces the equation of motion
 \be\label{cl}
 m_{\rm b}^{*\,2}\phi_{\rm cl}\simeq 2g n_{\rm f}^{(\pm)} [m_{\rm f}^{\phi}],
  \ee
where we indicated that the fermion density $n_{\rm f}^{(\pm)}$
depends on the classical field through $m_{\rm f}^{\phi}$. The
effective fermion mass (\ref{effermm}) then  becomes
 \be m_{\rm f}^{*} (T) =m_{\rm f}
-2\sqrt{J}-2g^2 n_{\rm f}^{(\pm)} [m_{\rm f}^{\phi}]/m_{\rm
b}^{*\,2}, \quad 2\sqrt{J}\ll m_{\rm f}.
 \ee
The total pressure is
 \be \label{prtot11} P_{\rm
tot}&\simeq&\frac{\pi^2 T^4}{90}-\frac{m_{\rm b}^2
T^2}{24}+2Tn_{\rm f}^{(\pm)} [m_{\rm f}^{\phi}]\nonumber\\
 &-&\frac{2g^2(n_{\rm f}^{(\pm)} [m_{\rm
f}^{\phi}])^2}{m_{\rm b}^{*\,2}} ,
 \ee
compare with Eq. (\ref{prtot}). The last term in (\ref{prtot11})
is  the contribution of the classical field $\delta P_{\rm cl}$.
For all temperatures $T\lsim T_{\rm bl}$ the classical field
contribution remains small, as consequence of the smallness of the
fermion density, but it drastically increases with decrease of the
effective boson mass at higher temperatures.

Note that similar consideration  can be performed in the chiral
$\sigma$ model, where in addition to the Yukawa interaction there
is the boson self-interaction, and already at zero temperature
there exists classical field $\phi_{\rm cl}\neq 0$.

\subsection{Hot Bose condensation}

For $T\gsim T_{\rm bl}$ typical fermion momenta $\sqrt{2m_{\rm
f}T}$ continue to remain much smaller than the mass $m_{\rm f}$.
However non-relativistic approximation for fermions that we have
used fails since typical deviation of the fermion energy from the
mass-shell becomes comparable with $m_{\rm f}$.  Nevertheless let
us extrapolate our results to higher temperatures. From
(\ref{bosc}) we see that at $\sqrt{J}(T_{\rm cB})=4g^2 n_{\rm
f}^{(\pm)}/m_{\rm b}^2$  squared of the effective boson mass
reaches zero and it becomes negative for $T>T_{\rm cB}$. Supposing
that $T_{\rm cB}$ deviates only a little from the value $T_{\rm
bl}^{\rm n.rel}=\sqrt{3}T_{\rm bl}$ (when $m_{\rm f}^{*}=0$) we
easily find
 \be
T_{\rm cB}\simeq T_{\rm bl}^{\rm n.rel}\left[1-\frac{T_{\rm
bl}^{\rm n.rel}}{m_{\rm f}}\mbox{ln}\frac{3^{3/2}\cdot 8m_{\rm
f}^2}{\pi^2 g m_{\rm b}^2}\right].
 \ee
 In Ref.
\cite{V04} in realistic relativistic framework  we found that
$T_{\rm cB}$ is in the vicinity of the value $T_{\rm bl}$, whereas
within non-relativistic model we find a larger value for the
critical density $T_{\rm cB}\simeq T_{\rm bl}^{\rm n.rel}$ (more
precisely $T_{\rm cB}$ is slightly below $T_{\rm bl}^{\rm n.rel}$
for $1\ll g\lsim 10$). This overestimation of the value $T_{\rm
cB}$ is a price paid  for simplicity of our non-relativistic
consideration of heavy fermions here.

Now we may consider a possibility of the
 {\em{ hot Bose condensation}} for $T>T_{\rm cB}$, see \cite{V04} where such a possibility
has been studied in the framework of relativistic model. In the
model with additionally introduced boson self-interaction $L_{\rm
int}=-\lambda \phi^4/4$ (with the coupling $\lambda >0$) the
condensate field is determined by minimization of the pressure in
the classical field variable. We find
 \be\label{cl1}
m^{*2}_{\rm b} \phi +\lambda  \phi^3 \simeq 2g n_{\rm f}^{(\pm)}
[m_{\rm f}^{\phi}] .
 \ee
Let us assume that for $T\gsim T_{\rm cB}$
 the contribution
$g n_{\rm f}^{(\pm)} [m_{\rm f}^{\phi}]$ is small and can be
neglected. It is so for $n_{\rm f}^{(\pm)} [m_{\rm f}^{\phi}]\ll
|m_{\rm b}^{*}|^3 /(g\sqrt{\lambda})$. Then, we may neglect  a
small classical field,  existing already for $m^{*2}_{\rm b}>0$,
see (\ref{cl}). A new condensate solution arises then for
$T>T_{\rm cB}$, when $m^{*2}_{\rm b}<0$. One may say that for $T=
T_{\rm cB}$ there occurs second order phase transition leading to
the hot Bose condensation. Inclusion into consideration of the
term $g n_{\rm f}^{(\pm)} [m_{\rm f}^{\phi}]$ results in a jump in
the value of the classical field at $T$ near $T_{\rm cB}$. It
means that actually the hot Bose condensation appears as the first
order phase transition. However, if the value of the jump is
small, one can neglect it and consider the second order phase
transition. Note that  saturation of the condensate field arises
for $T>T_{\rm cB}$ only due to the repulsive boson-boson
interaction ($\lambda
>0$).

Boson excitation spectrum is then reconstructed  and becomes
 \be
\om_{\rm b}^2 \simeq -2 m^{*2}_{\rm b}+p_{\rm b}^2
 \ee
(for $\alpha \ll 1$). The pressure is given by
 \be \label{prtot1}
&&P_{\rm tot}\simeq\frac{\pi^2 T^4}{90}+\frac{m_{\rm b}^2
T^2}{12}-\frac{g^2 n_{\rm
f}^{(\pm)}T^2}{2\sqrt{J}}+\frac{(4g^2n_{\rm
f}^{(\pm)}/\sqrt{J}-m_b^2)^2}{4\lambda}\nonumber\\ &&+2Tn_{\rm
f}^{(\pm)} .
 \ee
We see that the pressure acquires a positive contribution $\propto
1/\lambda$, large for small $\lambda$.

One could expect an anomalous enhancement of the boson (e.g. pion
and kaon) production at low momenta ($p_{\rm b}\lsim T$)  and an
anomalous behavior  of fluctuations, e.g. at LHC conditions, as a
signature of the hot Bose condensation for $T>T_{\rm cB}$, if a
similar phenomenon occurred in a realistic problem including all
relevant particle species.

\section{Concluding remarks}\label{Concluding}

We derived  exact and simplified expressions for thermodynamic
characteristics of the  matter of interacting particles with
finite widths. First, we disregarded  interaction terms and
considered system of "free resonances" using simplified ansatze
for spectral functions. Such a consideration might be  helpful in
description of dilute systems. We have shown that the model is
thermodynamically consistent. Our expressions demonstrate
deficiency of the use of approximation of constant particle width.
Such an approximation is often done to simplify the treatment of
the problem. Although at a higher density our ansatz for the
spectral function is definitely not valid and interaction and
potential energies are not small, the thermodynamical consistency
relations continue to be fulfilled.

Then we recovered  interaction terms and found simple exact
equations for thermodynamic values expressed in terms of spectral
functions.
 In the quasiparticle approximation these
expressions transform to the standard quasiparticle expressions.
For broad resonances and blurred particles all information on
medium effects is hidden in the form of  spectral functions. This
circumstance allows to model behavior of different systems
selecting relevant forms for spectral functions and checking that
consistency conditions are approximately fulfilled.

On examples of one-component non-relativistic system interacting
by paired potential and the relativistic boson gas with $\phi^4$
interaction, we have shown explicitly that the "free resonance"
term becomes dominating in the low density limit, provided the
width is finite but rather small.

We have shown, that in case of a fermion-boson interacting system,
the interaction contribution to thermodynamic quantities can be
expressed either in terms of the fermion spectral function or in
terms of the boson spectral function. Similar conclusion is done
for a system of two-flavor fermions interacting with a boson. Here
only one species is affected by the interaction, whereas two other
species represent free resonances or blurred particles. This
observation opens a possibility of approximate description of
multi-component systems, provided some species can be described in
the quasiparticle approximation and other species represent broad
resonances or blurred particles. The interaction energy can be
transported to the quasiparticle sub-systems. In the low density
limit the interaction part ceases more rapidly compared to a free
particle contribution. Then we deal with non-interacting
quasiparticles and free broad resonances.

As an illustrative example, we studied behavior of hot strongly
interacting ($g\gsim 1$) heavy fermion -- light boson sub-systems
at zero chemical potentials of species. Here in a broad
temperature interval ($T_{\rm bl}>T\gsim m_{\rm b}^* (T)$) bosons
can be well described in the quasiparticle approximation, whereas
fermions become blurred particles.

The following remark is in order. Although we considered the case
$m_{\rm b}\ll m_{\rm f}$, our results are more general, being
valid provided $m_{\rm b}^{*}\ll m_{\rm f}^{*}$. Thus even if
condition  $m_{\rm b}\ll m_{\rm f}$ is not fulfilled, with
increase of the temperature the system may jump to the regime
where blurred fermions undergo the Brownian motion in the bath of
effectively much less massive bosons, see \cite{V04} for more
details.

Fermion  particle distributions are significantly increased
compared to the standard Boltzmann distributions. The system
represents a gas mixture of boson quasiparticles interacting with
fermion-antifermion blurs. In thermodynamical values interaction
terms partially compensate each other. Thereby,   for very strong
coupling between species ($g\gg 1$) thermodynamical quantities of
the system, like the energy, pressure and entropy, prove to be
such as for the quasi-ideal gas mixture of quasi-free fermion
blurs and quasi-free bosons.

The latter observation gives us a hint for construction of
equation of state which might be valid at LHC conditions in some
temperature interval (below deconfinement temperature). One may
conjecture that the pressure is approximately presented as the sum
of all possible contributions of quasi-free bosons and quasi-free
baryon blurs. The latter contributions differ from the free baryon
particle terms only by pre-factors $F_{\rm f}(T)$, with subscript
$"{\rm f}"$ running over the baryon and antibaryon species. The
particle density for the given baryon species $n_{\rm f}$ is
enhanced compared to the corresponding Boltzmann value by the
factor $F_{\rm f}(T)$. The boson distribution depends on its
effective mass which may essentially decrease for the lightest
boson species. The ratio of the baryon (antibaryon) to boson
densities is expected to be significantly enhanced compared to the
ratio of the standard Boltzmann density for the given baryon
species to the massless boson density.

For $T>T_{\rm cB}$  there may arise a hot Bose condensation for
some boson species. If $T>T_{\rm cB}$ retained at freeze out
conditions, Bose condensation could manifest itself in an
anomalous enhancement of the corresponding boson  production at
low momenta (for the $s$-wave condensation discussed here).

Concluding, our treatment of particles with finite widths
(resonances and blurred particles)  is helpful in cases, when
explicit expressions for the spectral functions can be
constructed. E.g., description of a system of heavy fermions
strongly interacting with light bosons  at zero chemical
potentials of components represents such an example, cf.
\cite{V04}. Another example is the low density system. Then a
simplified treatment of "free resonances" with simple ansatze
spectral functions can be especially helpful provided  the broad
resonance appears, as the result of the interaction with other
particle species, which can be described within the quasiparticle
approximation.

\vspace*{5mm} {\bf Acknowledgements} \vspace*{5mm}

I am very grateful to Y.B. Ivanov, A.S. Khvorostuhin, J. Knoll,
E.E.~Kolomeitsev, V.V. Skokov and V.D. Toneev for  illuminating
discussions and valuable remarks. This work was supported in part
by the Deutsche Forschungsgemeinschaft DFG project 436 RUS
113/558/0-3, and the Russian Foundation for Basic Research RFBR
grant 06-02-04001.


\end{document}